# Effective Passivation of Exfoliated Black Phosphorus Transistors against Ambient Degradation

Joshua D. Wood[1†], Spencer A. Wells[1†], Deep Jariwala[1,2], Kan-Sheng Chen[1], EunKyung Cho[1], Vinod K. Sangwan[1,2], Xiaolong Liu[3], Lincoln J. Lauhon[1], Tobin J. Marks[1,2], and Mark C. Hersam[1,2,3*]

*[1]Dept. of Materials Science and Engineering, Northwestern University, Evanston, IL 60208*
*[2]Dept. of Chemistry, Northwestern University, Evanston, IL 60208*
*[3]Graduate Program in Applied Physics, Northwestern University, Evanston, IL 60208*

**Abstract**

Unencapsulated, exfoliated black phosphorus (BP) flakes are found to chemically degrade upon exposure to ambient conditions. Atomic force microscopy, electrostatic force microscopy, transmission electron microscopy, X-ray photoelectron spectroscopy, and Fourier transform infrared spectroscopy are employed to characterize the structure and chemistry of the degradation process, suggesting that $O_2$ saturated $H_2O$ irreversibly reacts with BP to form oxidized phosphorus species. This interpretation is further supported by the observation that BP degradation occurs more rapidly on hydrophobic octadecyltrichlorosilane self-assembled monolayers and on H-Si(111), versus hydrophilic $SiO_2$. For unencapsulated BP field-effect transistors, the ambient degradation causes large increases in threshold voltage after 6 hours in ambient, followed by a ~$10^3$ decrease in FET current on/off ratio and mobility after 48 hours. Atomic layer deposited $AlO_x$ overlayers effectively suppress ambient degradation, allowing encapsulated BP FETs to maintain high on/off ratios of ~$10^3$ and mobilities of ~100 $cm^2V^{-1}s^{-1}$ for over two weeks in ambient. This work shows that the ambient degradation of BP can be managed effectively when the flakes are sufficiently passivated. In turn, our strategy for enhancing BP environmental stability will accelerate efforts to implement BP in electronic and optoelectronic applications.

[*] Correspondence should be addressed to m-hersam@northwestern.edu

[†] These authors contributed equally.

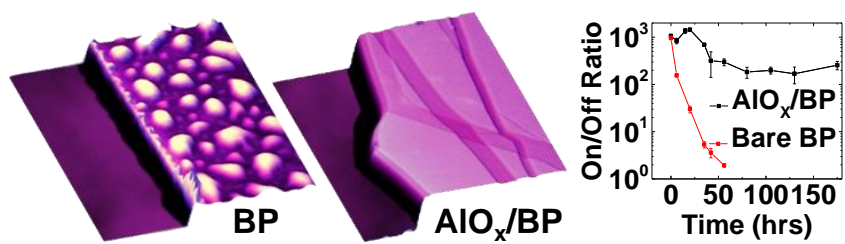

**TOC Figure**

**Manuscript**

Bulk black phosphorus (BP) is a layered, thermodynamically stable phosphorus allotrope[1, 2] with a graphitic structure. It is a bulk semiconductor with a band gap of ~0.3 eV[3, 4] and high carrier mobilities of ~1000 cm$^2$/V-s at room temperature.[5] Additionally, it exhibits anisotropic optical properties[6, 7] and superconducting characteristics,[8] while its layered structure enables use as an anode in Li-ion batteries.[9] Mechanical exfoliation of BP crystals has led to the isolation of few-layer and monolayer phosphorene flakes,[5, 6, 10-15] intensifying research on this two-dimensional (2D) nanomaterial. Exfoliated, p-type semiconducting BP flakes possess mobilities of ~200-1000 cm$^2$/V-s,[5, 6, 14] current on/off ratios of ~10$^4$-10$^5$,[5, 14] and anisotropic transport.[6, 16] Consequently, BP shows promise as a nanomaterial that could complement or exceed the electronic, spintronic, and optoelectronic properties of graphene. Nevertheless, questions remain regarding the nature of defects, contacts,[17] doping,[3] band transport,[7] and the chemical stability of BP and phosphorene.

Although graphene is chemically inert, other 2D nanomaterials such as graphane,[18, 19] fluorinated graphene,[20] and silicene[21] possess lower chemical stability. The reduced chemical stability of these alternative 2D nanomaterials is related to the energetics needed to maintain stable bonding configurations, which is affected by electrostatics and structural buckling.[21] Thus, ambient stability is likely to be a concern for BP since the phosphorus atoms have free lone pairs[22] and valence bond angles of 102°.[23] Early work involving ambient scanning tunneling microscopy (STM) measurements on bulk BP revealed the formation of pits and bubbles,[24] presumably arising from electrochemical reactions between the STM tip meniscus and BP crystal. While this induced chemical degradation of bulk BP provides reason for concern, the reactivity of thinner, exfoliated BP flakes in ambient conditions and the effect of chemical reactions on device-relevant electronic properties are completely unknown.

Towards this end, we examine here how exfoliated BP degrades to oxygenated phosphorus compounds in ambient environments through a comprehensive suite of microscopy and spectroscopy techniques including atomic force microscopy (AFM), electrostatic force microscopy (EFM), transmission electron microscopy (TEM), X-ray photoelectron spectroscopy (XPS), and Fourier transform infrared (FTIR) spectroscopy. Additionally, we investigate how BP field-effect transistors (FETs) degrade following ambient exposure, and then show that atomic layer deposition (ALD) of AlO$_x$ overlayers is an effective, scalable strategy for passivating BP flakes and FETs from ambient deterioration.

Figure 1 shows the structure of the present exfoliated BP flakes and FET devices (see Supporting Information for experimental details). Fig. 1A shows schematics for unencapsulated and ALD AlO$_x$ encapsulated BP devices. Few-layer BP flakes are exfoliated on 300 nm thick SiO$_2$/Si wafers and then located by optical contrast.[25] Flakes used in BP devices are less than 10 nm thick, and all other characterized flakes are less than 150 nm thick. An as-fabricated FET is highlighted in Fig. 1B, with no obvious optical evidence for degradation.[11] All flakes have the orthorhombic crystal structure characteristic of BP, as confirmed by Raman spectroscopy, high resolution transmission electron microscopy (TEM) imaging, and selected area electron diffraction (SAED) data (Fig. S1). After exfoliation in ambient, all flakes are stored at a nominal relative humidity of 36.3 ± 8.1%. Additional details of the ambient exposure conditions are given in the Supporting Information. Fig. 1C provides an AFM height image of a BP flake (~9.0 nm thick) shortly after exfoliation, showing small, topographic protrusions (hereafter termed "bubbles") above the BP flake plane. Fig. 1D shows this flake after one day of ambient exposure. For this flake, both the bubble density and the height of the bubbles relative to the flake plane have increased. In Figs. 1E and 1F, respectively, AFM height images are provided for the same BP flake after 2 and 3 days in ambient. On increased ambient exposure, the bubble density eventually decreases, evolving into wider and taller bubbles. These bubbles occur in BP, regardless of flake thickness (Fig. S2).

Figure S3 demonstrates AFM phase images for the different BP flakes in Fig. S2. While Figs. S3A, S3C, and S3D, respectively, show two regions of phase contrast, Fig. S2B shows three distinct regions of phase contrast. The two phase regions in Fig. S3A and others correspond to BP features in the height images. However, the third region in Fig. S3B does not correlate with the BP topography. AFM phase images can elucidate differences in tip-sample energy dissipation that are correlated with local changes in material properties.[26] Therefore, the distinct regions of local phase contrast highlight structural or chemical changes in the BP flake, suggesting that Fig. S3B reveals an intermediate chemical transformation in the BP with increasing ambient exposure. Moreover, with increasing ambient exposure, the roughness of the BP flakes on hydrophilic SiO$_2$ increases and is concentrated at the bubbles (Fig. S4). BP flake roughness also increases for flakes exfoliated on hydrophobic substrates such as octadecyltrichlorosilane (OTS) self-assembled monolayers on SiO$_2$ (Fig. S5). The overall flake roughness distribution increases monotonically with time (Fig.

S6), consistent with an earlier report[13] and indicative of volumetric expansion. All of these factors suggest that structural or chemical changes are occurring in the BP flakes upon ambient exposure.

In Fig. 2, we therefore use X-ray photoelectron spectroscopy (XPS) and Fourier transform infrared (FTIR) spectroscopy to assess whether chemical modifications, such as the formation of additional chemical bonds or a change in oxidation state, occur in BP upon ambient exposure. Fig. 2A shows P 2p core level XPS spectra of as-exfoliated BP flakes on SiO$_2$ for 0 hrs, 13 hrs, 1, day, 2 days, and 3 days, respectively, of ambient exposure. All spectra are calibrated to the binding energy of adventitious carbon (284.8 eV), and electrostatic charging is compensated using an Ar$^+$ flood gun (see Supporting Information for details). At 0 hrs of ambient exposure (black spectrum in Fig. 2A), the exfoliated BP exhibits a single spin-orbit split doublet at ~130 eV, consistent with previous XPS measurements on BP bulk crystals.[27, 28] Note that these spectra do not match those for red phosphorus (~129.8 eV), white phosphorus, or amorphous P–H.[27] A broad, s photoelectron Si satellite from the substrate 300 nm SiO$_2$ appears at ~126.5 eV. After 13 hrs of ambient exposure (maroon spectrum), the full-width at half-maximum (FWHM) for the BP increases, characteristic of some loss of long range order. After 1, 2, and 3 days in ambient (green, navy, and gray spectra, respectively), an additional doublet appears at ~134 eV. This feature is best assigned to phosphate species,[9, 29] although many oxidized phosphorus compounds exhibit peaks near ~134-135 eV.[30, 31] The large FWHM of the oxidized phosphorus peaks suggests BP amorphization, with multiple P–O and P=O bonding states. Thus, we assign the peaks to PO$_x$, for oxidized phosphorus species.[30]

Ambient species, particularly H$_2$O and O$_2$, are known to promote PO$_x$ formation in BP crystals.[24, 30, 32]. As such, we investigate the possibility that BP degradation might be suppressed by changing the wetting character of the underlying substrate, hence changing the surface diffusion rate of oxygenated H$_2$O. In Fig. 2B, substrate effects on BP degradation are assayed by examining OTS applied to hydrophilic SiO$_2$/Si (H$_2$O contact angle $\theta \sim 18°$), yielding coated, hydrophobic SiO$_2$/Si ($\theta \sim 91°$) (see Table S1). The starting XPS data for no ambient exposure are shown in black and pink for SiO$_2$/Si and OTS/SiO$_2$, respectively. After only one day of ambient exposure, the OTS/SiO$_2$ spectrum reveals significant BP degradation, with broadening of the BP P 2p doublet and the appearance of PO$_x$ peaks. Comparatively, the spectrum for hydrophilic SiO$_2$ after one day in ambient shows no significant PO$_x$ contribution or amorphization of BP. These spectra imply that the hydrophobic OTS/SiO$_2$/Si accelerates PO$_x$ formation. We also show this acceleration for core level spectra taken at the same time intervals as Fig. 2A (see Fig. S7). Examining the area under the $PO_x$ peaks relative to the BP peaks allows us to ascertain that $PO_x$ formation on OTS/$SiO_2$ is two times faster than on $SiO_2$ (Fig. S7B). We also note that optical imaging (Fig. S8) and statistical roughness data (Figs. S6B and S6D) confirm this accelerated degradation for BP on OTS/$SiO_2$ versus $SiO_2$.

Fourier transform infrared (FTIR) spectra further corroborate the presence of $PO_x$ species in degraded BP. Figure 2C shows FTIR spectra for bare $SiO_2$/Si (black) and BP on $SiO_2$/Si as exfoliated (maroon), 1 day (green), 2 days (navy), and over 1 week (gray) in ambient. In this spectral region, there are no $SiO_2$ phonons,[33, 34] so only vibrational modes related to confined,[35] phosphoric acid esters[36] (P–O stretch) are present. This P–O stretch is centered about ~880 cm$^{-1}$ and becomes more significant with further ambient exposure. By deconvolving the asymmetric Si–O TO and LO phonon contributions[33, 34] about ~1200 cm$^{-1}$, we can also observe phosphoryl (P=O) stretching modes[37] for the ambient exposed BP (Fig. S9). The coexistence of P–O and P=O stretching modes is consistent with BP degrading to $PO_x$ derivatives. Fig. 2D reveals FTIR data in the same spectral region for BP on hydrophobic OTS/$SiO_2$ for the time scales given in Fig. 2C. Like $SiO_2$, BP on OTS/$SiO_2$ exhibits P–O stretching modes. Spectral analysis about ~1200 cm$^{-1}$ (Fig. S9C) also shows P=O modes, emphasizing the evolution of BP to $PO_x$ during ambient degradation on OTS/$SiO_2$.

Figure 3 examines AFM phase images for BP on $SiO_2$ ($\theta \sim 18°$), OTS/$SiO_2$ ($\theta \sim 91°$), and H-passivated Si(111) ($\theta \sim 95°$) to better understand the degradation of BP on substrates of differing wetting character. Figs. 3A and 3B, respectively, show phase images for an as-exfoliated BP flake on $SiO_2$ (Fig. 3A) and after 1 day in ambient (Fig. 3B). Large bubbles, consistent with those reported in Fig. 1, appear after 1 day. Figs. 3C and 3D, respectively, provide phase images for as-exfoliated BP on OTS/$SiO_2$ (Fig. 3C) and after 1 day in ambient (Fig. 3D). For the as-exfoliated case on OTS/$SiO_2$, a higher density of bubbles appears relative to the hydrophilic $SiO_2$ control (Figs. 3A and 3B). After 1 day on OTS/$SiO_2$, the BP degradation is severe, with bubbles consuming the BP flake. From RMS roughness analysis (Fig. S6), the BP degradation on hydrophobic OTS is found to be 11 times faster than degradation on $SiO_2$ (same analysis as Figs. S4 and S5). As noted above, $PO_x$ formation for BP on OTS/$SiO_2$ is detected after 1 day in ambient (Figs. 2B and 2D), in accord with the rapid degradation observed in Figs. 3C and 3D. In the AFM phase images of Figs. 3E and 3F, we look at BP exfoliated on another hydrophobic substrate, H-Si(111).

Similar to BP on OTS/SiO$_2$, 1 day of ambient exposure for BP on H-Si(111) evidences severe degradation. In addition, large bubbles on H-Si(111) appear after only 2 hrs in ambient (Fig. S10).

To eliminate flake thickness or crystallographic effects on the BP degradation, we examine the same region of a BP flake on OTS/SiO$_2$ (Fig. S11), similar to the data presented for BP on SiO$_2$ in Figs. 1C-F. This flake demonstrates significant degradation bubbles after 1 day in ambient and is severely degraded after 3 days. Indeed, we observe movement and coarsening of BP degradation bubbles for BP on OTS/SiO$_2$ after only 15 min (Fig. S12). Conversely, BP on SiO$_2$ shows no movement in the same time frame. Thus, ambient BP degradation is accelerated on hydrophobic substrates versus hydrophilic ones. This difference in degradation kinetics can likely be attributed to the fact that oxygenated H$_2$O diffuses more rapidly on hydrophobic surfaces than hydrophilic surfaces.[38] Consequently, the oxygenated H$_2$O encounters the hydrophilic BP flakes more often on a hydrophobic surface than on a hydrophilic surface, ultimately leading to increased adsorption and/or intercalation of oxygenated H$_2$O, which then drives BP degradation.

To further illuminate the role of oxygenated H$_2$O in ambient BP degradation, BP flakes were characterized by AFM in an environmental cell. Specifically, when the environmental cell is filled with dry N$_2$, no bubbles or other evidence of BP degradation are detected in AFM height images after 4, 5, 15, and 24 hrs following exfoliation (Fig. S13). On another BP flake with no previous ambient exposure, humid air (~40% relative humidity) is introduced into the environmental cell. In Fig. S14 and in the supporting movie SM1, the growth and ripening of BP degradation bubbles is observed with increasing H$_2$O exposure time. Similar results are observed by TEM, as seen in Fig. S15 and supporting movie SM2. Therein, trapped bubbles are observed to diffuse and coalesce in a few seconds under the influence of the TEM electron beam. Since the TEM measurements are performed in ultrahigh vacuum (~10$^{-9}$ Torr), these bubbles are likely intercalated within the BP microstructure, suggesting that ambient degradation can occurs both within the BP crystal and on the surface. Also, TEM images (Fig. S16) and movies (supporting movies SM3 and SM4) on dry BP control samples show no modifications under the TEM beam, which implies that the changes in Fig. S15 and SM2 originate from encapsulated degradation species within the BP.

Figure 4 reveals the time dependence of BP FET ambient degradation. Specifically, Figs. 4A, 4B, and 4C show optical images of an as-fabricated BP FET and FETs after 1 and 3 days in ambient, respectively. Degradation bubbles appear after 1 day, consistent with the optical images in another report.[11] Moreover, the AFM height images of Figs. 4D-F provide additional evidence of bubble evolution in the BP flake. Note that the present FET processing conditions do not introduce any signature of $PO_x$ formation in the FETs (Fig. S17). BP degradation bubbles, like those seen in Figs. 1 and 3, appear after 1 day. In addition, Fig. 4F shows that the bubbles coalesce after three days in ambient. The bubbles do not appear under the electrical contacts to the BP FET, suggesting that BP can be passivated from ambient degradation by appropriately chosen coatings.

In addition to the widespread use of passivation in the semiconductor industry, other devices including carbon nanotube[39] and graphene FETs[40, 41] exhibit improved performance following encapsulation, further motivating an exploration of passivation coatings for BP FETs. Since previous reports have shown that atomic layer deposition (ALD)-derived dielectrics are compatible with BP FETs,[14, 42] we investigate ALD-derived $AlO_x$ as an encapsulation layer. Figures 4G and 4H show optical images of a ~30 nm thick $AlO_x$ encapsulated BP flake after 1 and 7 days in ambient, respectively. Compared to Figs. 4B and 4C, no BP degradation related bubbles appear for ambient-exposed, $AlO_x$-capped BP. Furthermore, Figs. 4I-L show AFM height images for the same region of the flake in Fig. 4G after 1, 2, 3, and 7 days in ambient. This BP flake shows no degradation bubbles, despite the prolonged ambient exposure. BP flakes of differing thickness are also robustly passivated against degradation, withstanding even 34 days in ambient (Fig. S18). The $AlO_x$ encapsulated BP flakes show no increase in RMS roughness (Fig. S6C and S6D) with time. Finally, the flakes are chemically pristine, as they possess no signatures of $PO_x$ formation, shown by the purple IR spectrum in Fig. 2C.

To explore the passivating effects of ALD-derived $AlO_x$ on FET electrical properties, BP devices were fabricated by electron beam lithography with electrodes consisting of ~2 nm Ti and ~70 nm Au. The liftoff procedure was performed using dry acetone in a glove box to minimize degradation during processing. An as-fabricated BP FET transfer curve is shown in Fig. 5A, and yields an initial $I_{ON}/I_{OFF}$ ratio > $10^3$ with a field-effect hole mobility of 74 cm$^2$ V$^{-1}$ s$^{-1}$, comparable to literature values for exfoliated few-layer BP FETs.[10] After fabrication and initial measurement, the devices were stored in ambient conditions in the dark (see Supporting Information for details), allowing the effects of ambient exposure on device performance to be probed. Within 6 hrs of ambient exposure, the FET transfer curve shifts significantly, with the $I_{OFF}$ current increasing by a factor of ~7 and the threshold voltage increasing by 22 V. These observations are attributed to p-type doping from atmospheric adsorbates, similar to effects observed with carbon nanotubes[43] and graphene.[40] This trend of increased doping continues up to 20 hrs of ambient exposure. However, by 35 hrs, this threshold voltage shift slows, and the transfer curve qualitatively changes, concurrent with the $PO_x$ formation seen spectroscopically (Fig. 2). Note that the degradation timescale is accelerated versus that in Fig. 2A-B, as the 0 hr case here corresponds to the completion of device fabrication, rather than exfoliation. Using electrostatic force microscopy (EFM) on a grounded Au pad, the BP degradation bubbles are found to possess compromised electrical conductivity relative to the rest of a representative BP flake (Fig. S19). These results are consistent with a severe decrease in FET conduction following 35 and 56 hrs of ambient exposure.

An identically prepared BP FET was electrically characterized immediately after fabrication and then after encapsulation with a 30 nm thick film of ALD $AlO_x$. The ALD encapsulation results in lower hysteresis and an increase in the $I_{ON}/I_{OFF}$ ratio from 285 to over $10^3$ (Fig. 5B). The encapsulated device also displays an initial hole mobility of 46 $cm^2V^{-1}s^{-1}$ and linear output characteristics (Fig. S20). The encapsulated device was then stored under identical ambient conditions as the unencapsulated device. During the first 20 hrs of exposure, the threshold voltage increases slightly, but the device performance is otherwise unchanged (Fig. 5C). We observe even better performance for $AlO_x$ encapsulated BP FETs with Ni/Au contacts (Fig. 5D). After 20 hrs, the $I_{ON}/I_{OFF}$ ratio decreases slightly, and the threshold voltage continues to increase, although this process is retarded significantly relative to the unencapsulated device. Indeed, the $AlO_x$ encapsulated BP FET is still functioning after one week of ambient exposure (175 hrs) with an $I_{ON}/I_{OFF}$ ratio of over ~250 and a mobility of ~44 $cm^2V^{-1}s^{-1}$. For convenience, a log scale plot of the FET transfer curves are provided in Fig. S21.

Comparing two Ti/Au devices, the $I_{ON}/I_{OFF}$ ratio of the unencapsulated BP device falls rapidly upon ambient exposure. Within 6 hrs, the $I_{ON}/I_{OFF}$ ratio drops by a factor of 7 in the unencapsulated device, while the encapsulated device is unchanged (Fig. 5E). Within 35 hrs, the $I_{ON}/I_{OFF}$ ratio of the unencapsulated device has dropped below 10, rendering it unusable as a switch. However, the $AlO_x$ encapsulated device has an $I_{ON}/I_{OFF}$ ratio that is effectively unchanged and remains greater than $10^3$ for the first 20 hrs, with the ratio remaining between 180 and 300 through the remainder of the 175 hrs investigated. Device mobility (Fig. 5F), while only declining slightly in the unencapsulated device during the first 20 hrs, drops precipitously afterwards. After 36 hrs of ambient exposure, the mobility falls by more than two orders of magnitude, from an initial value of 74 to 0.25 $cm^2V^{-1}s^{-1}$. Conversely, ALD-derived $AlO_x$ encapsulation preserves the mobility above 40 $cm^2V^{-1}s^{-1}$ for the entire 175 hrs investigated.

Upon further investigation, more BP FETs were fabricated and encapsulated under ALD $AlO_x$, but were fabricated using 20 nm Ni and 60 nm Au electrodes rather than the previous Ti/Au contacts. As observed in the previous devices, ALD encapsulation again lowers device hysteresis and threshold voltage, while increasing the $I_{ON}/I_{OFF}$ ratio and hole mobility (Fig. S22). In these Ni/Au devices, the transfer curves change more slowly than in the encapsulated Ti/Au devices, indicating that contact metals may also play a role in device performance (Fig. 5D). These 10 encapsulated Ni/Au devices began measurement with an average $I_{ON}/I_{OFF}$ ratio of ~7500 and a mobility of ~49 $cm^2V^{-1}s^{-1}$. After 2 weeks of ambient exposure, they displayed an average $I_{ON}/I_{OFF}$ ratio of ~3000 and mobility of ~53 $cm^2V^{-1}s^{-1}$ (Fig. S23).

As described recently, the buckled structure of BP produces lone pairs that can promote $O_2$ chemisorption.[44] While this theoretical report states that it is unfavorable for chemisorbed $O_2$ to dissociate and form P–O bonds, interstitial O could be introduced readily into BP.[44] In turn, the interstitial O will cause volumetric expansion[13] and provide hydrophilic dipoles[11] for $H_2O$ and other molecules to interact with BP. These theoretical results correlate well with the present observations of BP degradation and bubble expansion following exposure to ambient. In the presence of confined,[35, 45] oxygenated $H_2O$, the BP breakdown likely proceeds through hygroscopic $P_4O_6$ and $P_4O_{10}$ intermediaries,[24] ultimately producing the $PO_x$ species observed in XPS and FTIR (Fig. 2). Furthermore, the greater spin-orbit coupling – more than 0.84 eV – and binding energy downshift in the XPS spectra for ambient aged BP (Fig. 2) suggest some polymeric red phosphorus formation in the flakes.[27] In either case, covalent modification or transformation of BP by ambient exposure compromises electrical properties and degrades FET performance. We attribute the effectiveness of ALD-derived encapsulation layers in suppressing this degradation to the conformal nature of ALD thin film growth, which passivates both edges and top surfaces equivalently, regardless of aspect ratio.[46] Subsequently, both edge and top surface adsorbate access is occluded by ALD application, arresting ambient degradation mechanisms and significantly enhancing BP FET environmental stability.

In summary, the degradation of exfoliated BP following ambient exposure has been characterized structurally and spectroscopically. Ambient adsorbates irreversibly convert BP into $PO_x$ compounds, fundamentally altering BP electronic and material properties. Hydrophobic substrates such as OTS/SiO$_2$/Si and H-Si(111) are found to accelerate ambient degradation by promoting concentration of atmospheric adsorbates at BP flakes. In contrast, ALD-derived AlO$_x$ thin films effectively encapsulate BP, preserving high carrier mobilities and I$_{ON}$/I$_{OFF}$ ratios in BP FETs. By extensively characterizing ambient degradation mechanisms and delineating a scalable ALD mitigation strategy, the results of this study should accelerate future efforts to employ BP in applications of technological relevance.

**Acknowledgments**

This research was supported by the Materials Research Science and Engineering Center (MRSEC) of Northwestern University (NSF DMR-1121262), the Office of Naval Research (N00014-14-1-0669), and the Keck Foundation. The work made use of the NUANCE Center, which has received support from the MRSEC (NSF DMR-1121262), NSEC (NSF EEC-0118025/003), State of Illinois, and Northwestern University. Raman spectroscopy was conducted at Argonne National Laboratory at the Center for Nanoscale Materials, a U.S. Department of Energy, Office of Science, Office of Basic Energy Sciences User Facility, under Contract No. DE-AC02-06CH11357. S.A.W. acknowledges a National Defense Science and Engineering Graduate (NDSEG) Fellowship, and D.J. acknowledges an SPIE Education Scholarship. The authors thank J. McMorrow for the use of his relative humidity logging software.

**Supporting Information**

Discussion of the materials and methods used, additional experimental data, and experimental movies. This document is available free of charge via the Internet at http://pubs.acs.org.

**Author Contributions**

†These authors (J.D.W. and S.A.W.) contributed equally to this work.

J.D.W., S.A.W., D.J., K.-S.C., and X.L. exfoliated BP from bulk crystals. J.D.W. performed AFM, XPS, FTIR, Raman spectroscopy, and thin film transfer. S.A.W. performed AFM on BP devices. J.D.W. and D.J. took contact angle measurements. E.C. produced H-Si(111) substrates, and K.-S.C. produced OTS on SiO$_2$/Si substrates. K.-S.C. and J.D.W. gathered AFM data in an environmental cell. X.L. collected TEM and diffraction data. S.A.W. prepared devices by electron beam lithography. J.D.W., S.A.W., D.J., and K.-S.C. deposited thin film passivation layers by ALD. S.A.W., D.J., V.K.S., and J.D.W. measured charge transport in the devices with assistance from K.-S.C. All authors discussed results and participated in the preparation of the manuscript. L.J.L., T.J.M., and M.C.H. supervised the project.

**Note Added in Proof**

During the review process, we became aware of two related preprints[32, 47] on the ambient stability of BP.

**Conflict of Interest**

The authors declare no competing financial interest.

**Figures**

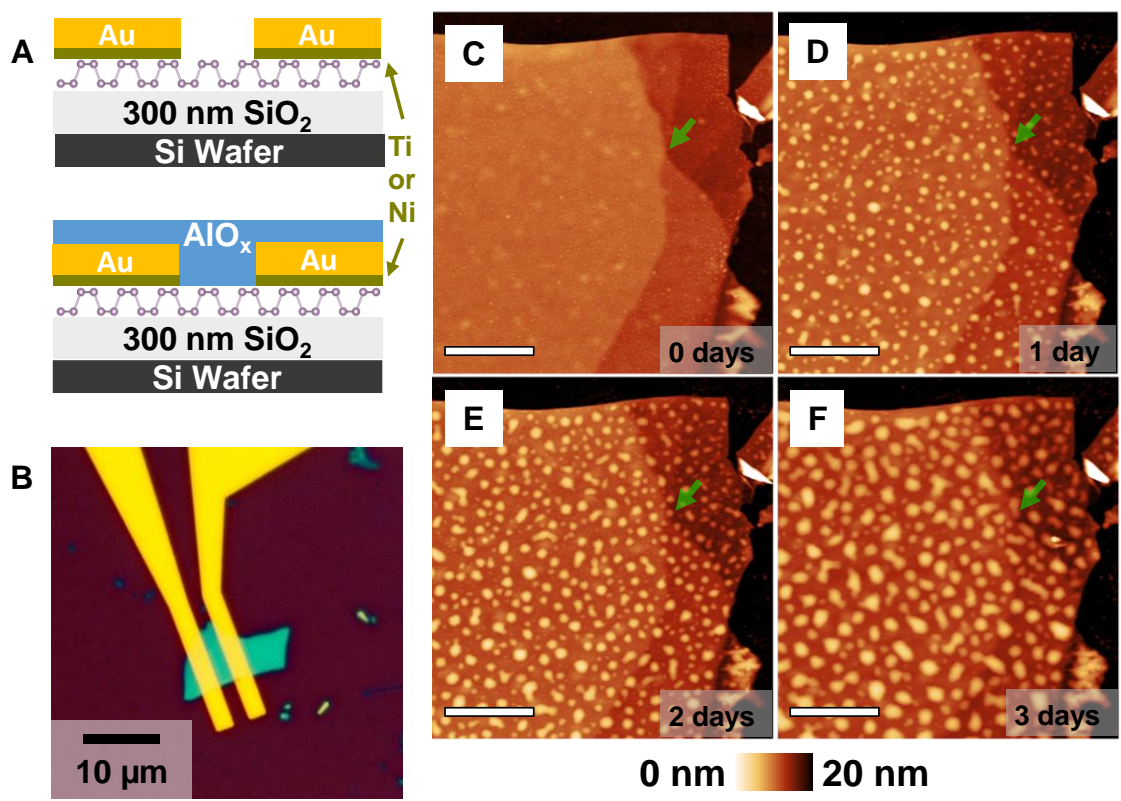

**Figure 1.** Time dependent degradation of exfoliated black phosphorus (BP) flakes on a SiO₂ substrate. **(a)** Schematics of unencapsulated and AlOₓ encapsulated BP field-effect transistors (FETs). **(b)** Optical image of a typical BP FET on 300 nm SiO₂. AFM height images for a 9.0 nm thick unencapsulated BP flake **(c)** after exfoliation, **(d)** 1 day in ambient conditions, **(e)** 2 days in ambient conditions, and **(f)** 3 days in ambient conditions. Green arrow shows the same region on the flake, and the scale bars are 1 μm.

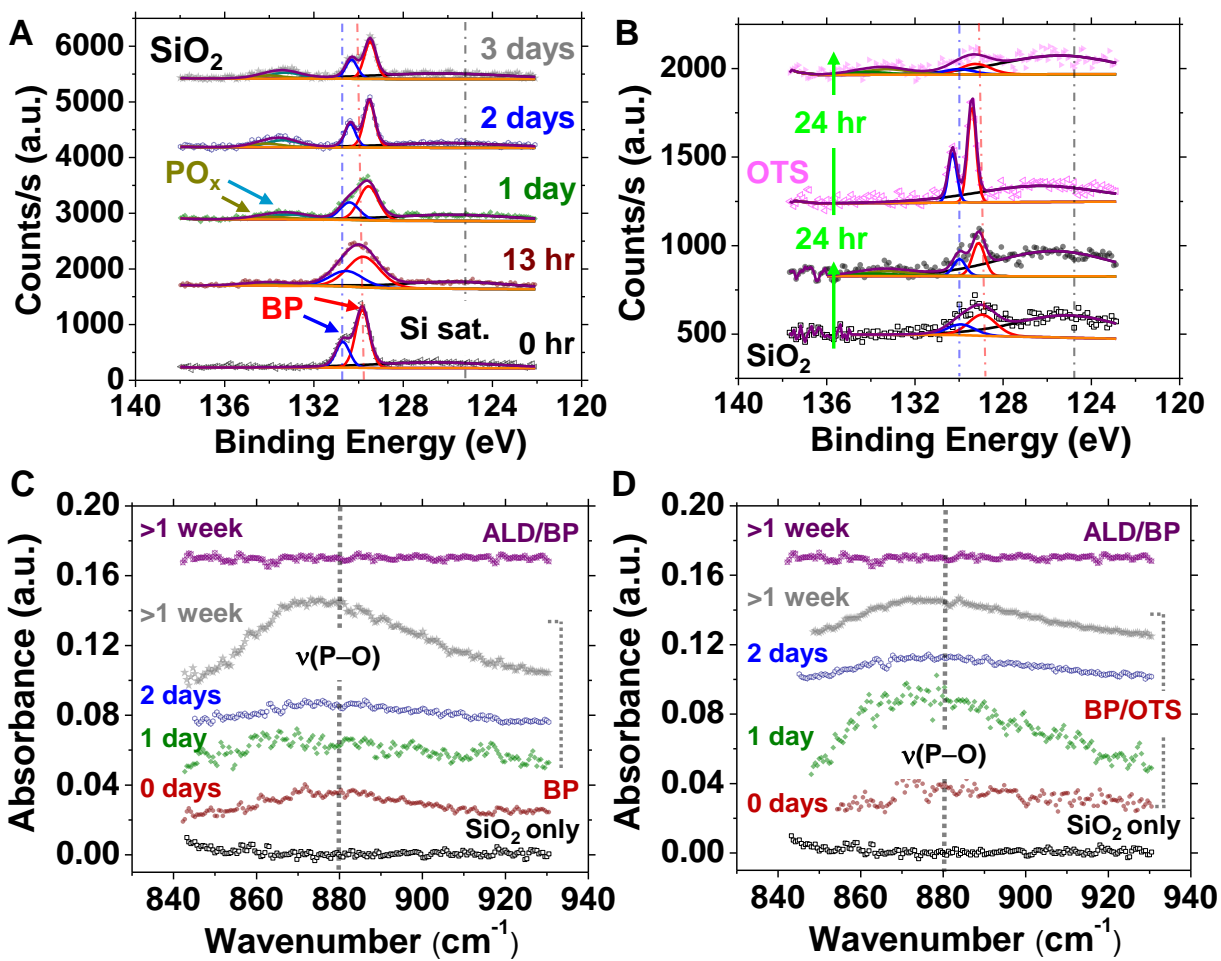

**Figure 2.** Irreversible chemical degradation of exfoliated BP to PO$_x$ derivatives. **(a)** P 2p core level X-ray photoelectron spectra (XPS) for BP at 0 hr, 13 hr, 1 day, 2 days, and 3 days ambient exposure. PO$_x$ peaks appear after 1 day. **(b)** P 2p core level XPS data for BP on hydrophilic SiO$_2$ and hydrophobic self-assembled OTS on SiO$_2$. Severe BP degradation and PO$_x$ peaks appear after 1 day for OTS/SiO$_2$, with the BP on SiO$_2$ control sustaining minor degradation in the same time. **(c)** Normalized Fourier transform infrared (FTIR) absorbance spectra for SiO$_2$, BP on SiO$_2$, ALD AlO$_x$ capped BP on SiO$_2$. Stretching modes characteristic of phosphoric acid esters (P–O) appear for on BP/SiO$_2$ with time but are absent on SiO$_2$ and AlO$_x$/BP. **(d)** Normalized IR absorbance spectra for BP on OTS/SiO$_2$ versus ambient exposure time. P–O stretching mode evolves with lower ambient exposure time for BP/OTS/SiO$_2$. All spectra are vertically offset for clarity.

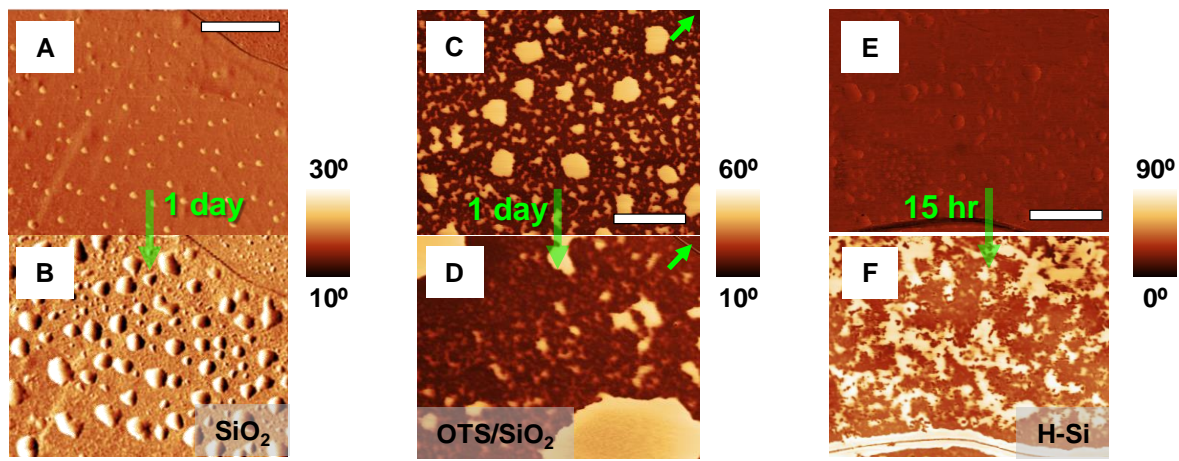

**Figure 3.** Acceleration of BP degradation on hydrophobic substrates. AFM phase images of a BP flake on hydrophilic SiO$_2$ ($\theta = 18°$) **(a)** after exfoliation and **(b)** 1 day later. AFM phase images of a BP flake on hydrophobic OTS/SiO$_2$ ($\theta = 91°$) **(c)** after exfoliation and **(d)** 1 day later. Green arrows denote the same location. BP degradation is more pronounced on hydrophobic OTS/SiO$_2$ versus hydrophilic SiO$_2$, implicating oxygenated H$_2$O as a degradation source. AFM phase images for the same BP region on hydrophobic H-Si(111) ($\theta = 95°$) **(e)** after exfoliation and **(f)** 15 hrs later. The BP flake on H-Si(111) has again suffered degradation in under one day. Scale bars are 1 μm.

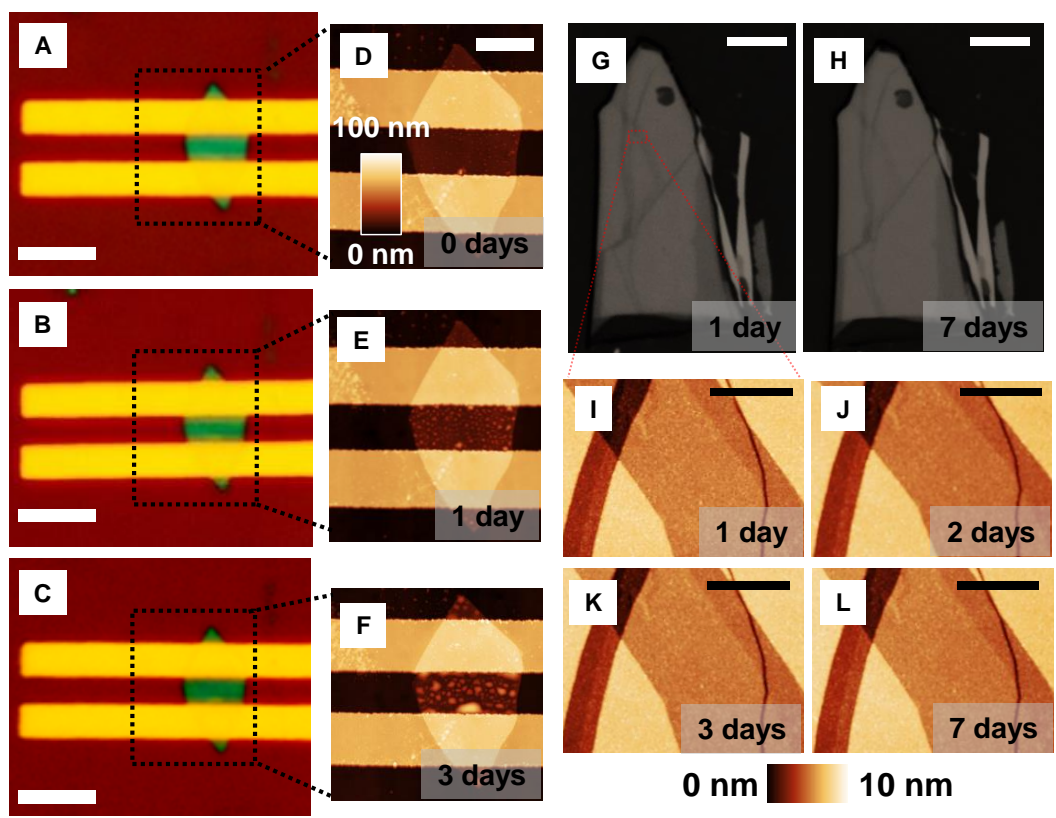

**Figure 4.** Passivation of exfoliated BP on $SiO_2$ by ALD-derived $AlO_x$. Optical images of an unencapsulated BP field-effect transistor (FET) **(a)** 0 days after fabrication, **(b)** 1 day in ambient, and **(c)** 3 days in ambient. Scale bars are 5 μm, and flake thickness is ~8.9 nm. Corresponding AFM height images of the boxed region after **(d)** 0 days, **(e)** 1 day, and **(f)** 3 days. Scale bars are 2 μm, and the height scale for (d-f) is shown inset on (d). BP degradation bubbles are apparent following ambient exposure. Optical images of an $AlO_x$ encapsulated BP flake (~92.7 nm) on $SiO_2$ after **(g)** 1 day and **(h)** 7 days in ambient. No degradation evident. Scale bars are 5 μm. AFM height images for the encapsulated BP in (g) after **(i)** 1 day, **(j)** 2 days, **(k)** 3 days, and **(l)** 7 days in ambient. No BP degradation occurs despite the extended ambient exposure. Scale bars are 500 nm.

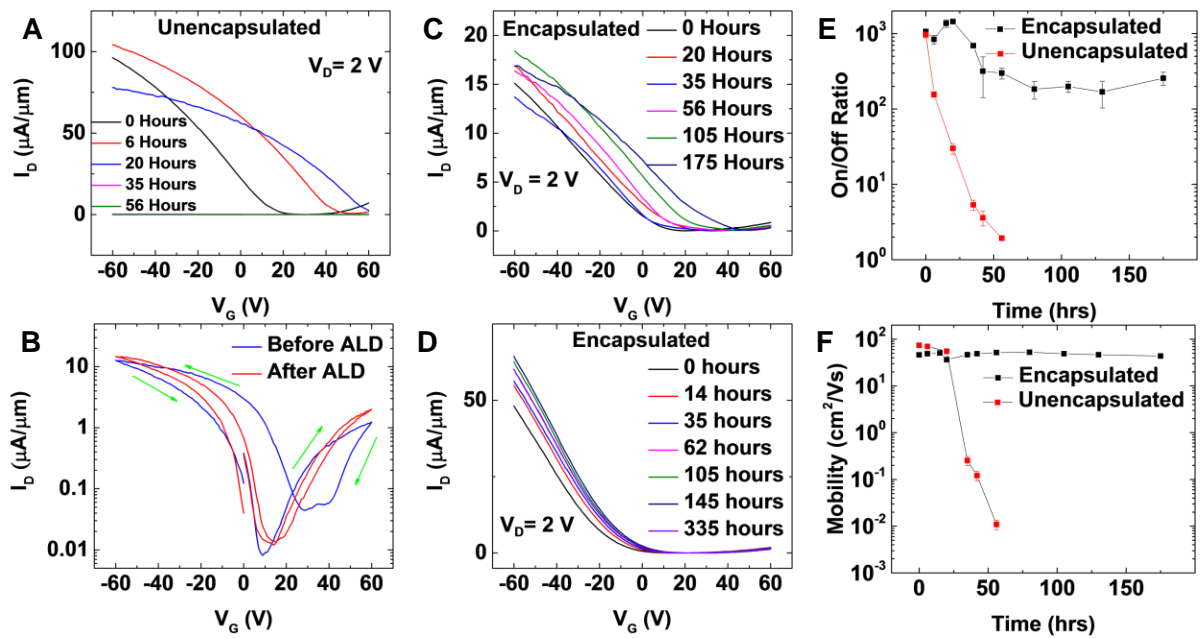

**Figure 5.** Time dependence of few-layer BP FET device characteristics. **(a)** Transfer curves for an unencapsulated BP FET with Ti/Au contacts, measured as a function of ambient exposure time. **(b)** Transfer curves for a BP FET measured immediately before and after encapsulation. **(c)** Transfer curves for a ~30 nm thick ALD AlO$_x$ encapsulated BP FET with Ti/Au contacts, measured as a function of ambient exposure time. **(d)** Transfer curves for a ~30 nm thick ALD AlO$_x$ encapsulated BP FET with Ni/Au contacts, measured against ambient exposure time. Comparison of the **(e)** I$_{ON}$/I$_{OFF}$ ratio and **(f)** hole mobility for encapsulated and unencapsulated BP FETs versus ambient exposure time.

*Supporting Information*

# Effective Passivation of Exfoliated Black Phosphorus Transistors against Ambient Degradation

Joshua D. Wood[1†], Spencer A. Wells[1†], Deep Jariwala[1,2], Kan-Sheng Chen[1], EunKyung Cho[1], Vinod K. Sangwan[1,2], Xiaolong Liu[3], Lincoln J. Lauhon[1], Tobin J. Marks[1,2], and Mark C. Hersam[1,2,3*]

[1]*Dept. of Materials Science and Engineering, Northwestern University, Evanston, IL 60208*
[2]*Dept. of Chemistry, Northwestern University, Evanston, IL 60208*
[3]*Graduate Program in Applied Physics, Northwestern University, Evanston, IL 60208*

**Contents:**

[*] Correspondence should be addressed to m-hersam@northwestern.edu
[†] These authors contributed equally.

**Materials and methods**

*Mechanical exfoliation*

Black phosphorus (BP) crystals are produced by a commercial supplier (Smart Elements). The BP source crystal is stored in the dark in a $N_2$ glove box. We exfoliate BP using standard Scotch tape. We remove air exposed pieces of the Scotch tape to provide a fresh piece of tape for exfoliation. We exfoliate onto 300 nm thick $SiO_2$ on degenerately doped Si, 300 nm thick $SiO_2$ on Si with shadow evaporated Cr/Au markers, OTS on $SiO_x$ (native oxide) on Si (as prepared below), and H-Si (as prepared below).

Unencapsulated BP flakes that were used in charge transport measurements were covered in 950 A4 poly(methyl methacrylate) (PMMA) (MicroChem, Inc.) within 10 min after exfoliation. The PMMA was coated at 4000 RPM for 45 seconds and baked out at ~180 $^\circ$C for 90 seconds. The flake of interest was identified by optical contrast and patterned by electron beam lithography.

*Sample exposure*

All prepared BP/substrate samples were either exfoliated immediately before measurement or otherwise stored in a dark $N_2$ glove box. When the samples were exposed to ambient conditions, they were kept in opaque sample containers in dark sample drawers. The samples were removed from their containers and measured in ambient, unless other conditions were required (e.g. ultra-high vacuum XPS measurements, device transport under vacuum, etc.). Measurements were performed with a relative humidity of $36.3 \pm 8.1\%$. Over the course of sample preparation, the relative humidity in the laboratory ranged from 16.8% to 57.4%, with a mode of 43.3%.

*Hydrogen passivation*

H-passivated Si(111) was produced from As-doped Si wafers (~0.002 $\Omega$·cm) with a native oxide (Virginia Semiconductor). The Si(111) substrate was degreased in acetone and isopropyl alcohol for 30 min each. The substrate was cleaned in piranha solution (70 mL sulfuric acid : 30 mL hydrogen peroxide) for 30 min and rinsed in ultrapure DI (18.2 M$\Omega$·cm) water. An H-terminated surface was prepared using ammonium fluoride ($NH_4F$). The $NH_4F$ solution was deoxygenated using Ar sparging for 30 min prior to use. The sample was soaked into the $NH_4F$ solution for 10 min, rinsed with DI water, and dried with nitrogen gas. The H-Si(111) was stored in the glove box until use.

*OTS on SiO$_x$/Si*

Octadecyltrichlorosilane (OTS) was purchased (Sigma-Aldrich) and used as received without further treatment. A 5 mM OTS solution is prepared by dissolving the OTS molecules in anhydrous toluene (Sigma-Aldrich). Before the formation of the OTS self-assembled monolayer, the SiO$_x$/Si wafer is cleaned by ultrasonication in acetone and isopropanol followed by thorough rinsing in distilled water. The wafer is then annealed on a hot plate at 150 °C for 10 min and subsequently submerged in the 5 mM OTS solution for 24 hrs to form a monolayer of OTS on SiO$_x$/Si.

*Transmission electron microscopy (TEM)*

BP flakes were exfoliated onto 300 nm of SiO$_2$/Si and then spin coated (3000 RPM, 60 s) with 950 A4 PMMA (MicroChem, Inc.). We removed the anisole from the PMMA with a bakeout at ~150 °C for 10 min. We exposed the PMMA/BP/SiO$_2$/Si samples to 2 M KOH overnight, which freed the PMMA/BP flakes from the SiO$_2$/Si. The PMMA/BP films were rinsed in ultrapure DI (18.2 MΩ·cm) for 5 min. We then picked up the PMMA/BP films with a Quantifoil (Au, 1.2 μm mesh, Ted Pella) TEM grid. We were careful to prevent rupture of the Quantifoil from the surface tension of the H$_2$O. The H$_2$O trapped at the interface of the PMMA/BP and the TEM grid was minimized by 10 min of ~70 °C and ~150 °C heating. Using a necked Erlenmeyer flask, we dissolved the PMMA through hot acetone vapor exposure. The grids were exposed to hot acetone for at least 8 hrs, and then they were stored in ambient until TEM measurement. The TEM measurement was performed with a JEOL JEM-2100 TEM at an accelerating voltage of 200 keV.

*X-ray photoelectron spectroscopy (XPS)*

XPS spectra were gathered in ultrahigh vacuum (UHV) at a base pressure of ~5×10$^{-10}$ Torr using a Thermo Scientific ESCALAB 250 Xi. The X-ray source was a monochromated Al K$_\alpha$ source at 1486.7 eV with a 400 μm spot size. The XPS spectra had a binding energy resolution of 0.1 eV, and a charge compensating flood gun was employed for all samples (e.g. BP/SiO$_2$, BP/H-Si, etc.). The core level spectra were collected at a pass energy of 10 eV and a 100 ms dwell time. Five scans were averaged for each core level. Even with charge compensation, all core levels were charge corrected to adventitious carbon at 284.8 eV. Using the software suite Avantage (Thermo Scientific), all subpeaks were fitted against a "Smart" background, which is a modified Shirley background [1]. All subpeaks had full-width at half-maximum (FWHM) greater than 0.5 eV, and each subpeak for a given core level shared the same amount of Gaussian-Lorentzian (GL) mixing.

If GL mixing became 0% (Lorentzian) or 100% (Gaussian), then the subpeaks were locked at 30% GL mixing, as is typically used in CasaXPS and other tools. The p core levels for phosphorus and silicon were fitted with doublets. Subpeaks were added until the residual level was minimized.

*Fourier transform infrared (FTIR) spectroscopy*

FTIR spectra were collected on a Thermo Nicolet Nexus 870 with the Tabletop optics module. Samples were placed on a Ge window (65° fixed incident angle) in an attenuated total reflectance (ATR) setup. $LN_2$ was used to cool the system prior to use, and the ATR module was continually purged with nitrogen. A nitrogen background with no sample was taken prior to any sample measurement, and the system collected data in absorbance mode. The nominal resolution was 0.5 cm$^{-1}$ for a 256 scan acquisition, using an aperture of 10. With the OMNIC software, absorbance data were corrected from effects due to the ATR background and baseline. Fityk [2] was employed to normalize the data to the intensity of the Si TO phonon (~1070 cm$^{-1}$). After normalization, the data were fitted with Voigt subbands about a splined baseline in the spectral region of interest. All spectra in Figs. 2C,D and Fig. S9 were vertically offset for clarity.

*Atomic force microscopy (AFM)*

All height and phase measurements were performed using ~300 kHz Si cantilevers in normal tapping mode on an Asylum Cypher AFM. Images were taken in the repulsive phase regime using at least 512 samples per line. The scanning rate was 1.5 Hz or lower.

AFM imaging employing $N_2$ flow was performed in an environmental cell attached to the Cypher ES scanner. Ultra-high purity grade $N_2$ was continuously flowed through the cell for 10 min before installing the BP sample. After the BP was exfoliated onto a substrate, it was immediately transferred to the environmental cell, with fewer than 10 seconds of exposure to ambient air. BP flakes of interest for AFM scanning were identified *in situ* with the built-in optical microscope in the Cypher system. During AFM scanning, $N_2$ was continuously flowed through the cell with the optical microscope light kept on.

Local roughness calculations were calculated using the first two nearest neighbors for each AFM height pixel. This roughness value was determined by the root-mean-squared (RMS) roughness calculation for those first two nearest neighbors (25 adjacent pixels total). For estimating kinetics of BP breakdown (see BP on OTS discussion in the main text) on hydrophobic and hydrophilic substrates, we assumed all roughness values greater than 1 nm to be representative of the onset of breakdown.

*Raman spectroscopy*

Raman spectra were taken on a Renishaw confocal Raman system using inVia software. Data were collected at 514 nm at ~0.1 mW through a 100X objective for 45 s, and Lorentzian line shapes were used to fit the data against a linear baseline in Fityk [2].

*Contact angle measurements*

We used contact angle measurements to determine the hydrophobic character of our as-prepared $SiO_2$, plasma-cleaned $SiO_2$, OTS/$SiO_x$/Si, and H-Si(111) samples. A droplet of $H_2O$ (~1-10 µL) was placed on the substrate in question, and the droplet was imaged at the solid-liquid interface by a digital camera. The digital image was analyzed by ImageJ (DropAnalysis plugin, LBADSA technique) [3], fitting a contact angle for each substrate. The results are the average of three trials for each substrate, given in table S1.

**Table S1. Contact angle measurements for the substrates used in exfoliation**

| Substrate | Contact Angle (deg) |
|---|---|
| Plasma-cleaned $SiO_2$ | < 10° |
| Degreased $SiO_2$ | 18° |
| OTS/$SiO_x$/Si | 91° |
| H-Si(111) | 95° |

*Electron beam lithography (EBL) and device fabrication*

Device features were defined using electron beam lithography (EBL) at a dose of ~360 µC/cm². The PMMA was developed in air with a 1:3 ratio of methyl isobutyl ketone (MIBK) to isopropyl alcohol (IPA). Following development, 5 nm of titanium and 50 nm of gold were evaporated to serve as alignment markers, and the PMMA was dissolved for 30 minutes in dry acetone in a glove box to liftoff the metal. To make contact to the flake, the same procedure was followed, using 2 nm of titanium and 70 nm of gold instead for the unencapsulated devices, and 20 nm nickel and 60 nm of gold for the encapsulated devices.

*Atomic layer deposition (ALD) of AlO$_x$*

ALD of AlO$_x$ was performed using trimethylaluminum (TMA) and H$_2$O as precursors in a Cambridge NanoTech reactor. For XPS and AFM studies, 3 nm of alumina was deposited at room temperature. During the ALD process, pulses of TMA precursor are introduced before those of H$_2$O, thereby minimizing adsorption and absorption of oxygenated H$_2$O into the BP. For BP FET passivation ~30 nm of alumina (AlO$_x$) was deposited. To form a seed layer and protect BP from potential damage due to high temperature exposure to oxygenated H$_2$O, the first ~3 nm were deposited at room temperature with prolonged purge time for both precursors [4]. This was followed by the remainder of the film deposited at 150 ºC with a normal purge time. No initial seeding layer such as metallic aluminum or any other material was used.

*Charge transport measurements*

Electrical characterization of BP FETs was performed on a Lakeshore CRX 4K under a pressure of ~10$^{-4}$ Torr, using two Keithley Source Meter 2400 to measure current-voltage characteristics. All electrical measurements were carried out at room temperature.

The mobility values in Fig. 5 and elsewhere were calculated using equation 1:

$$\mu_{eff} = \frac{L g_d}{W C_{ox} V_{DS}} \tag{1}$$

where $\mu_{eff}$ is the field effect mobility, $L$ is the channel length, $g_d$ is the transconductance, $W$ is the channel width, $C_{ox}$ is the oxide capacitance (11 nF·cm$^{-2}$ for 300 nm thick thermal SiO$_2$), and $V_{DS}$ is the source-drain voltage. Transconductance was calculated by taking the numerical derivative of the transfer curves, binning the nearest 5 values together to minimize the effect of measurement noise, and then taking the maximum binned slope.

Transfer curve data were collected at $V_{DS}$ = 0.005, 0.01, 0.05, 0.1, 0.5, 1, and 2 V. Data presented in Figs. 5A-D are from $V_{DS}$ = 2 V curves taken at different times. Error bars are defined as the standard deviation of the calculated mobility value at all values of $V_{DS}$ divided by the number of investigated $V_{DS}$ values at which the quantity was measured (i.e., standard error).

**Figures**

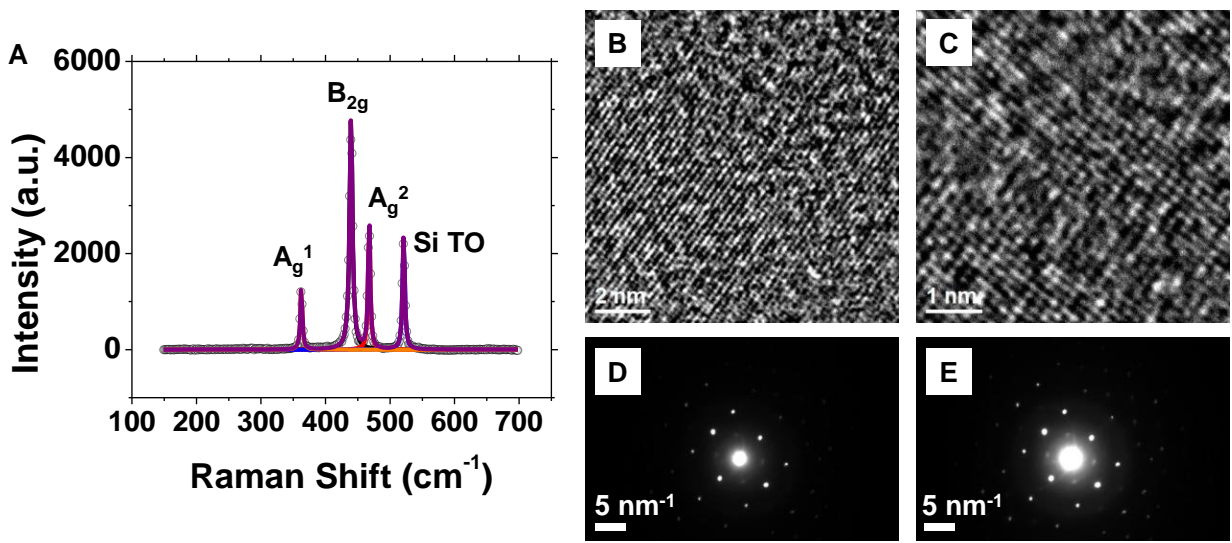

**Figure S1. General characterization of exfoliated black phosphorus (BP) flakes. (a)** Raman spectrum for BP on SiO$_2$. The three Raman bands (i.e., A$_g^1$, A$_g^2$, and B$_{2g}$) and their intensities versus the Si TO band (520 cm$^{-1}$) are consistent with BP (~100 nm thickness) on SiO$_2$ [5]. Bright field TEM images for transferred, exfoliated BP at **(b)** lower magnification and **(c)** high resolution. Expected lattice fringes for the BP surface appear. **(d, e)** Selected area electron diffraction (SAED) patterns for different regions of a BP flake stored under a N$_2$ glove box environment prior to imaging. Crystal symmetry matches orthorhombic BP. No ambient degradation is evident in the flake, as corroborated by supporting movies SM3 and SM4.

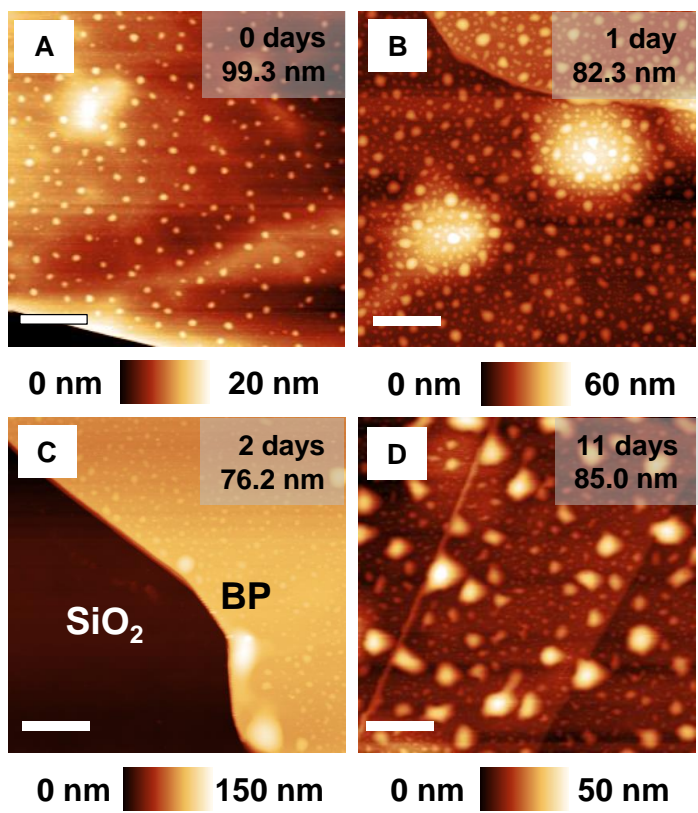

**Figure S2. BP degradation for flakes of differing thickness.** Height images for four BP flakes of differing thickness (thicknesses specified in the figure labels). **(a)** Partly degraded BP flake after exfoliation. **(b)** Another flake after 1 day in ambient. **(c)** A third flake after 2 days in ambient. **(d)** A fourth flake after 11 days in ambient. Scale bars are 1 µm.

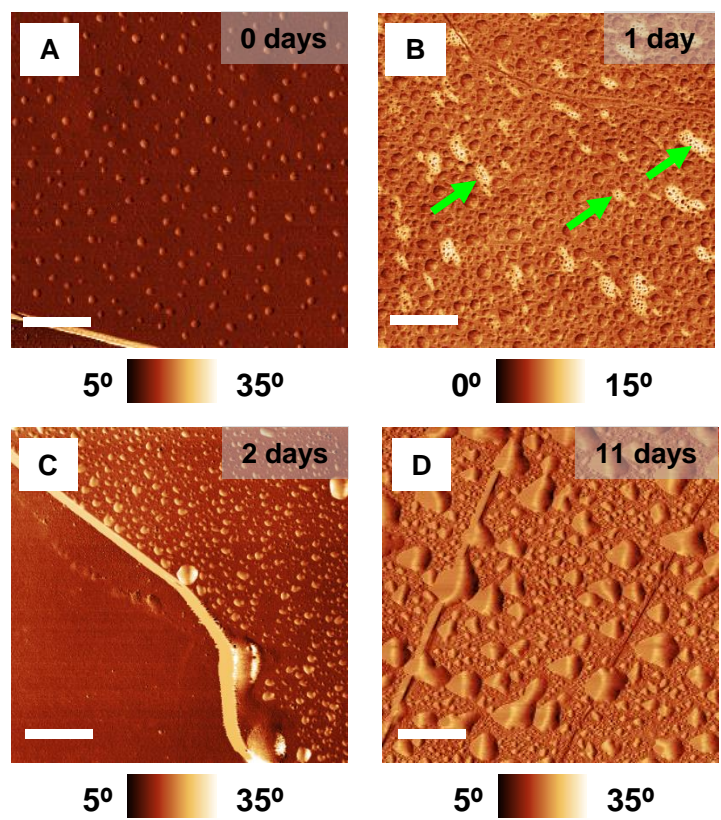

**Figure S3. AFM phase images showing BP degradation.** Phase images for the flakes in Fig. S2 of this document main manuscript (**a**) after exfoliation, (**b**) after 1 day in ambient, (**c**) after 2 days in ambient, and (**d**) after 11 days in ambient. The three different phase regions in (b) are consistent with an intermediate material transformation (green arrows) during BP degradation. Scale bars are 1 μm.

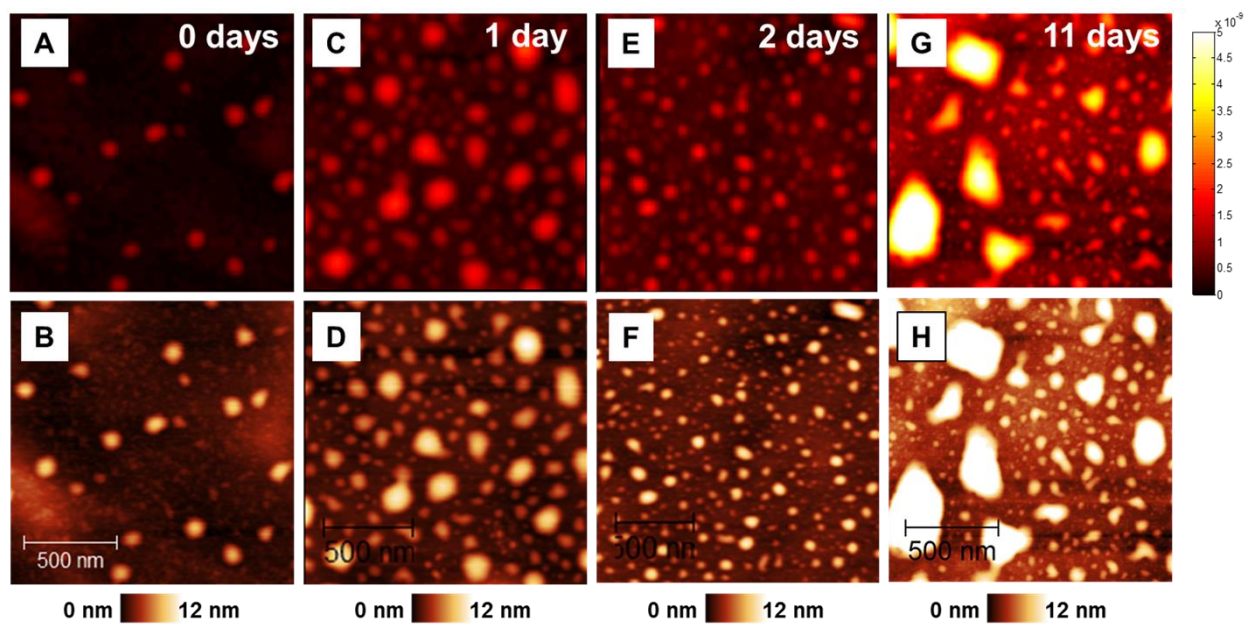

**Figure S4. Increasing BP roughness with time for different BP flakes. (a)** Local BP roughness for a flake after exfoliation on SiO$_2$. Roughness is determined by a RMS measurement of the two nearest neighbors at each pixel. **(b)** AFM height image for the area used in (a). **(c)** BP roughness after 1 day of ambient exposure. **(d)** AFM height image for (c). **(e)** BP roughness after 2 days in ambient. **(f)** AFM height image for (e). **(g)** BP roughness after 11 days in ambient. **(h)** AFM height image for (g). Roughness scale ranges from 0 to 5 nm.

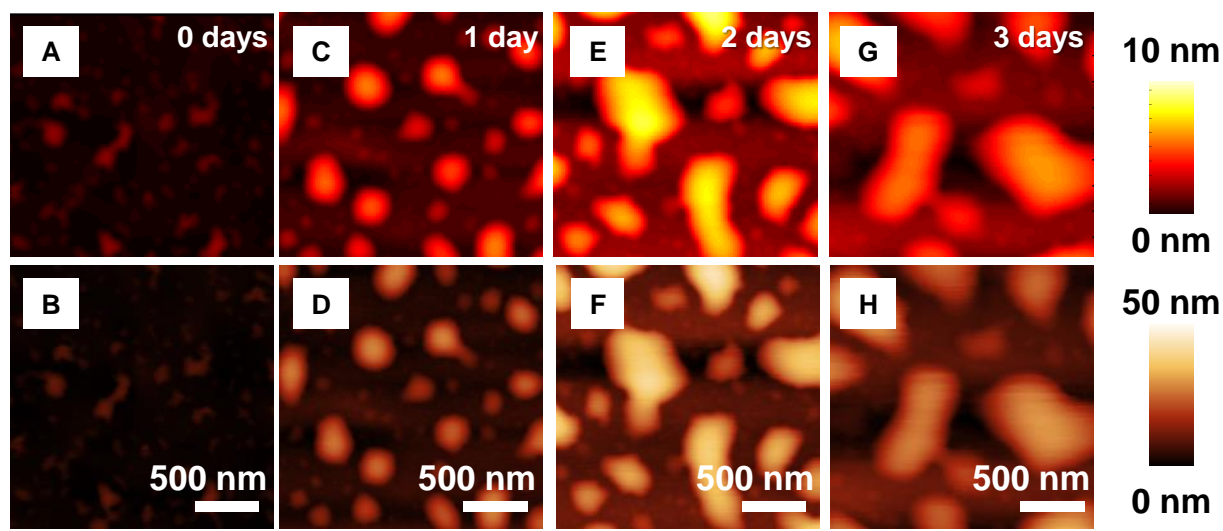

**Figure S5. Increasing BP roughness with time for the same flake on OTS. (a)** Local BP roughness for a flake (~77.6 nm thickness) on hydrophobic OTS/SiO$_2$. Roughness is calculated in the same manner as Fig. S4. **(b)** AFM height image for the area used in (a). BP roughness after **(c)** 1 day, **(e)** 2 days, and **(g)** 3 days of ambient exposure. **(d, f, h)** AFM height images for (c), (e), and (f), respectively. Flake roughness increases more quickly on hydrophobic OTS/SiO$_2$ than on hydrophilic SiO$_2$.

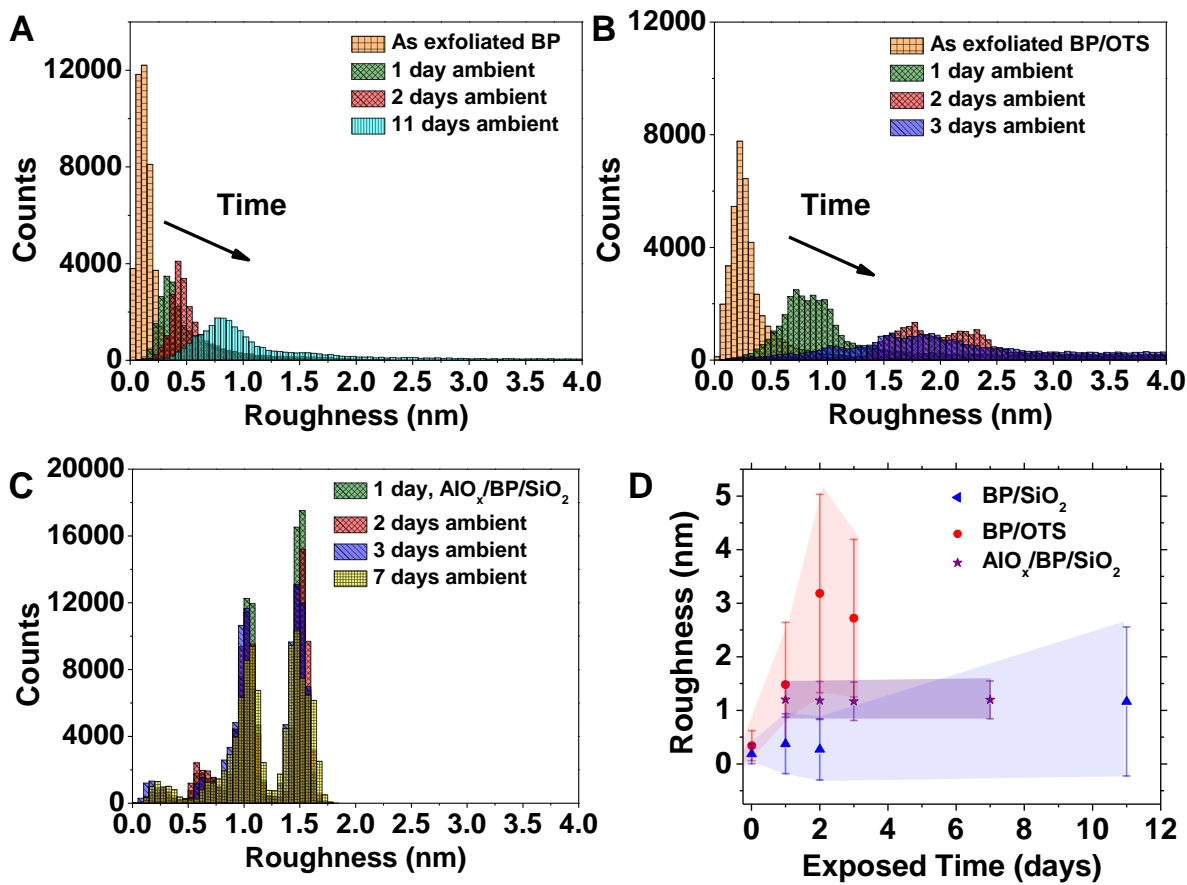

**Figure S6. Roughness statistics for BP on SiO₂, BP on OTS, and AlOₓ encapsulated BP. (a)** AFM data derived roughness histogram for BP on hydrophilic SiO₂. Roughness monotonically increases with ambient exposure. **(b)** Roughness histogram for BP on hydrophobic OTS/SiO₂. The roughness increases more rapidly for BP on OTS, highlighting the accelerated ambient BP degradation on hydrophobic substrates. **(c)** Roughness histogram for AlOₓ encapsulated BP on SiO₂. All roughness histograms overlap despite extended ambient exposure, demonstrating BP passivation from ambient degradation by the AlOₓ overlayer. **(d)** Evolution of roughness data with ambient exposure for unencapsulated BP/SiO₂, unencapsulated BP/OTS/SiO₂, and encapsulated AlOₓ/BP/SiO₂. Encapsulated BP flakes do not change with time, whereas BP on hydrophobic OTS/SiO₂ degrades ~11 times faster than on hydrophilic SiO₂. All flakes were measured at a nominal relative humidity of ~36%.

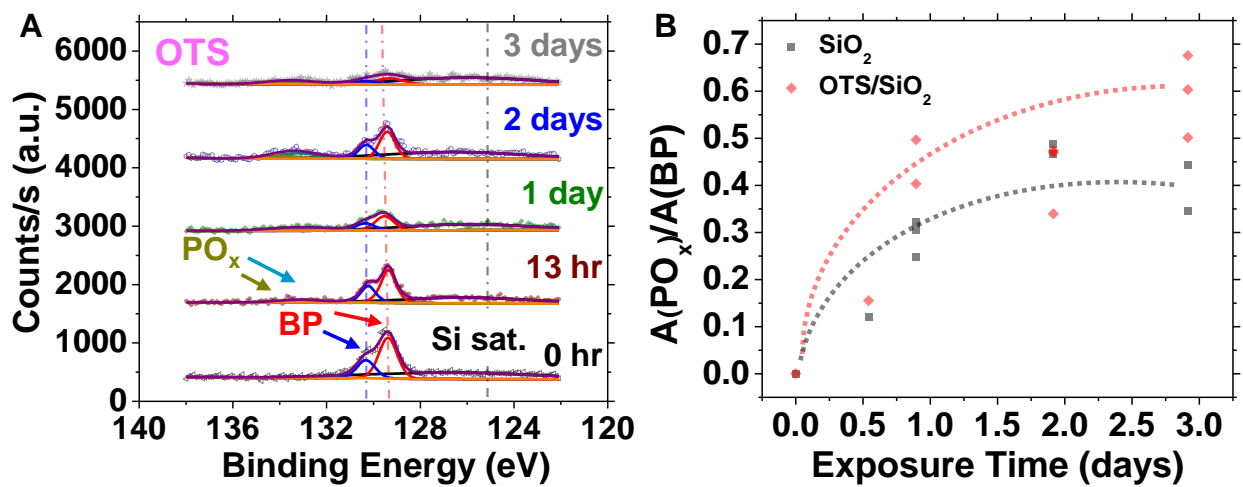

**Figure S7. P 2p XPS spectra for BP on SiO₂ and OTS versus exposure time. (a)** XPS P 2p core level data for BP/OTS/SiO$_2$ at different ambient exposure times. All spectra vertically offset for clarity. Oxidized phosphorus (PO$_x$) peaks appear after 13 hrs in ambient for OTS. After 3 days in ambient, the BP XPS signal is noisy and broad, indicative of a loss of long range order and substantial BP degradation. **(b)** XPS determined PO$_x$ percentage with respect to time for BP/SiO$_2$ and BP/OTS/SiO$_2$. Dotted lines are a guide to the eye. Percentage is determined as the area under the oxidized phosphorus (PO$_x$) subpeaks (~133.5 eV) divided by the area under the BP subpeaks (~130.0 eV). PO$_x$ percentage increases more rapidly for BP/OTS/SiO$_2$ versus BP/SiO$_2$, characteristic of faster BP degradation.

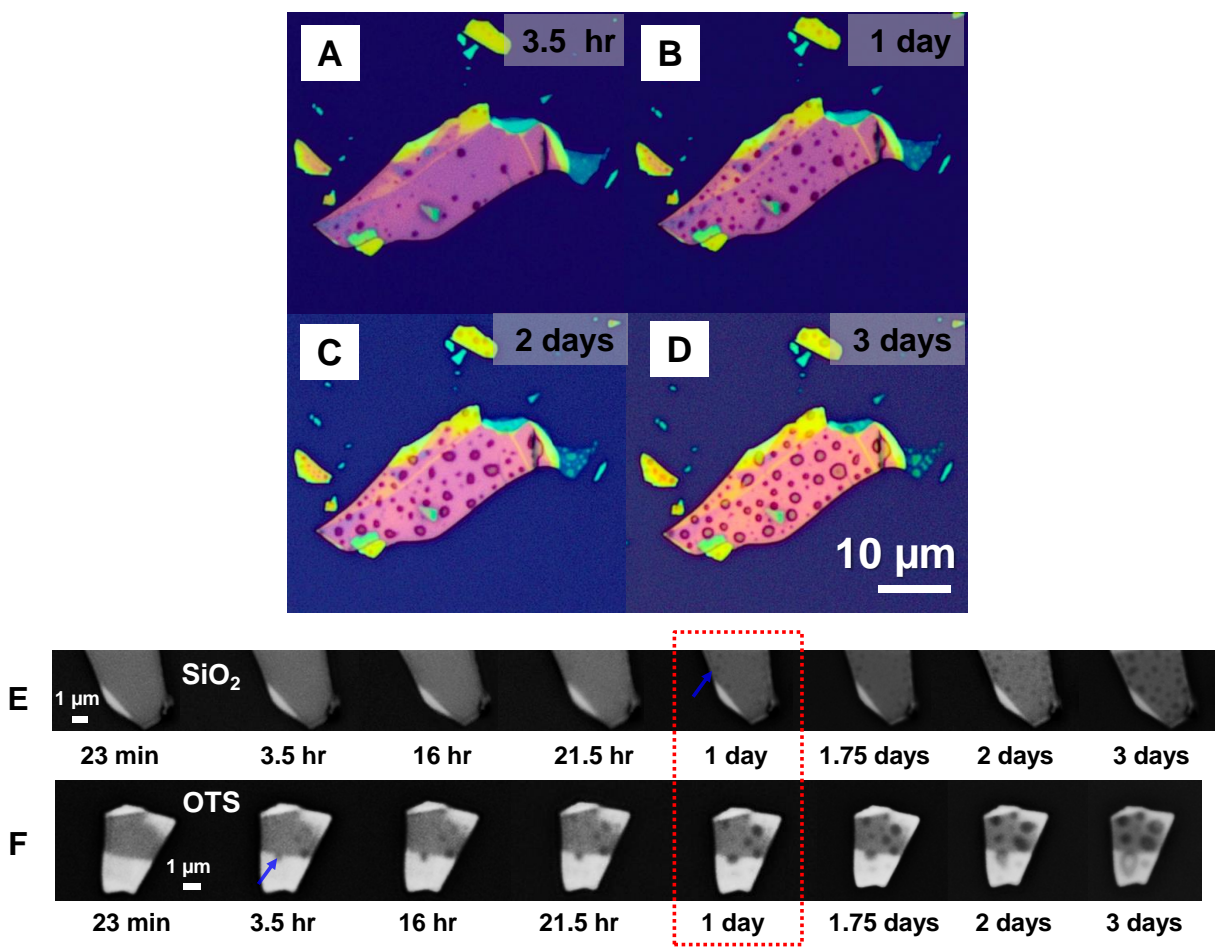

**Figure S8. Optical images of accelerated BP degradation on OTS/SiO₂ versus SiO₂.** Image of a large BP flake on OTS/SiO₂ after (**a**) 3.5 hrs, (**b**) 1 day, (**c**) 2 days, and (**d**) 3 days in ambient. BP degradation bubble formation is obvious. Flake integrity appears compromised after 1 day for BP/OTS/SiO₂. Time elapsed optical image montage for (**e**) BP/SiO₂ and (**f**) BP/OTS/SiO₂. Degradation appears after 3.5 hrs for BP/OTS/SiO₂. Conversely, it takes one day for optical degradation to appear on BP/SiO₂ (blue arrow). BP on both substrates is severely degraded after 3 days in ambient.

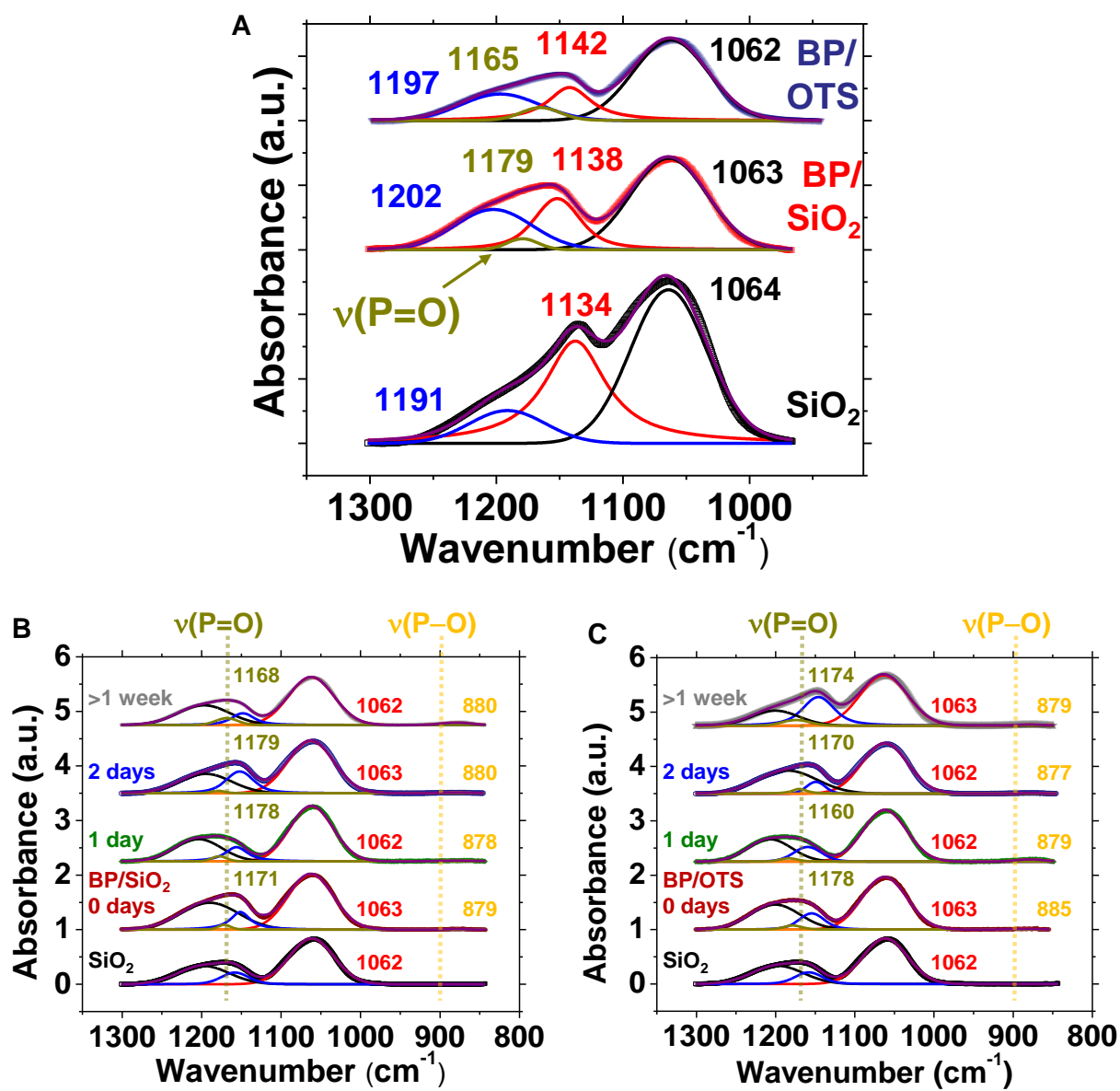

**Figure S9. Infrared (IR) spectra for BP on SiO₂ and OTS versus exposure time. (a)** Normalized IR spectra for a $SiO_2$ control, BP/$SiO_2$, and BP/OTS/$SiO_2$. Both BP/$SiO_2$ and BP/OTS/$SiO_2$ samples were exposed to ambient for more than 1 week. The exposed BP samples demonstrate a phosphoryl (P=O) stretching mode from the ambient degradation. Normalized IR spectra with respect to ambient exposure time for **(b)** BP/$SiO_2$ and **(c)** BP/OTS/$SiO_2$. As highlighted in Figs. 2C and 2D of the main manuscript, stretching modes for phosphoryl and phosphoric acid esters (P–O) are present in the BP/$SiO_2$ and BP/OTS/$SiO_2$ samples. Relative to $SiO_2$ only, the BP/OTS/$SiO_2$ samples have a more rapid appearance of the stretching modes. All spectra vertically offset for clarity.

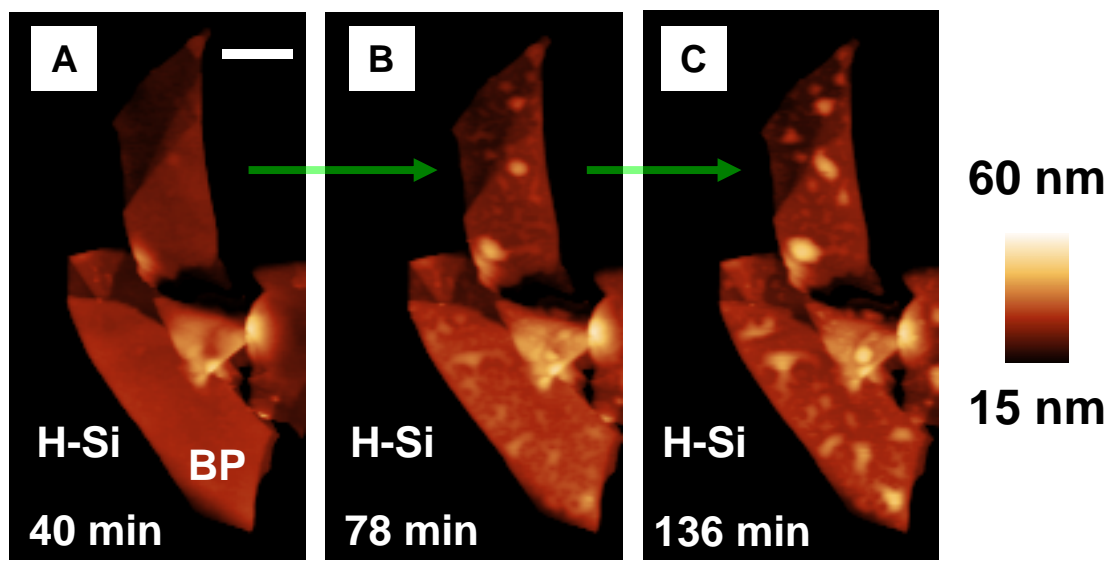

**Figure S10. Accelerated BP degradation on H-Si(111).** AFM height images for the same BP flake on H-Si(111) **(a)** 40 min after exfoliation, **(b)** 78 min after exfoliation, and **(c)** 136 min after exfoliation. Degradation bubbles under the BP appear in this short amount of time, likely from oxygenated $H_2O$ and/or BP decomposition products. All three images share the same plane fit and height scale, indicating that the apparent flake height is increasing with ambient exposure. Scale bar is 500 nm, which is the same length scale for (a-c).

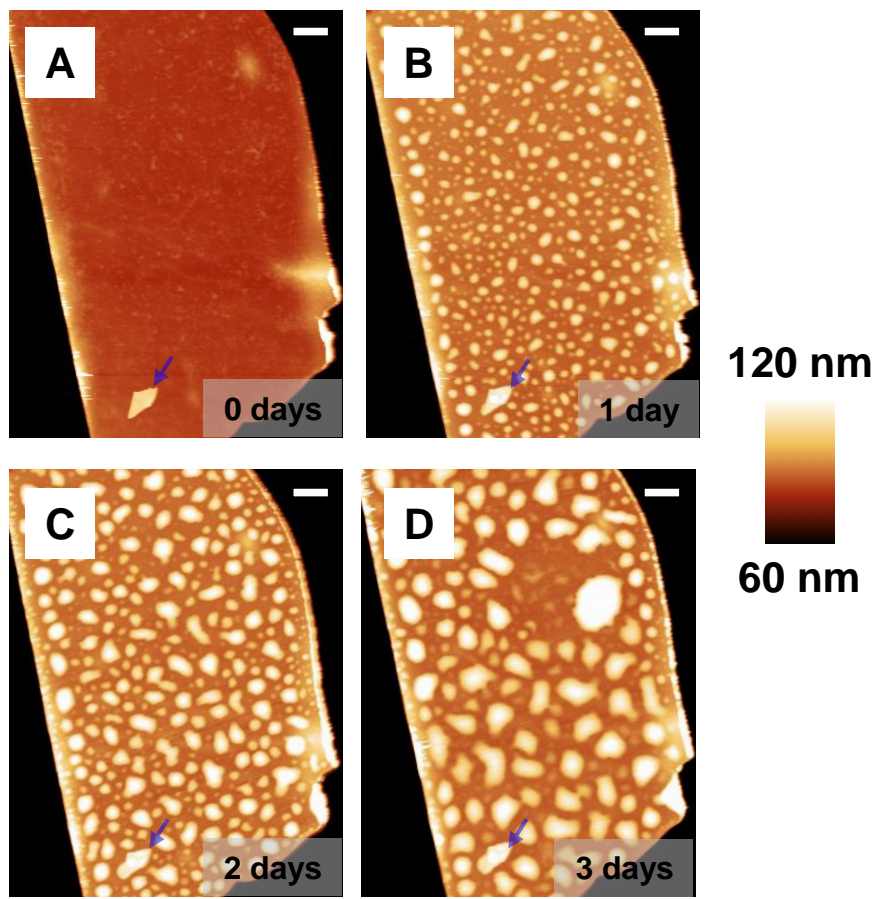

**Figure S11. Accelerated BP degradation on OTS/SiO₂.** AFM height images for the same ~76.8 nm BP flake on OTS/SiO$_2$ after **(a)** exfoliation, **(b)** 1 day in ambient, **(c)** 2 days in ambient, and **(d)** 3 days in ambient. Degradation bubbles appear more quickly for BP on OTS/SiO$_2$ versus SiO$_2$ alone (Figs. 1C-F). Flake height is increasing with ambient exposure time, due to oxygenated H$_2$O absorption in the BP. Scale bar is 1 µm, and the blue arrow highlights the same region of the flake.

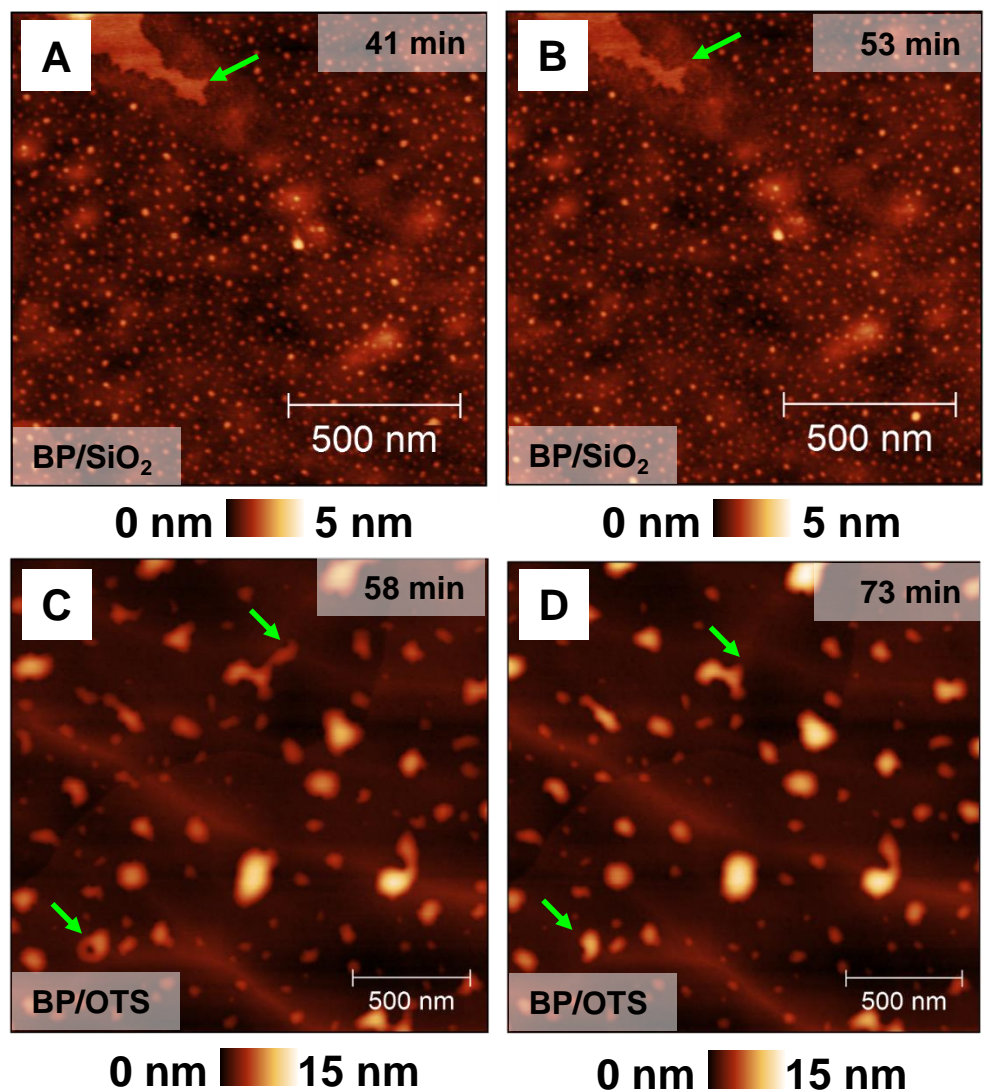

**Figure S12. Early degradation transients for BP/SiO₂ versus BP/OTS/SiO₂.** AFM height images for a ~9.0 nm BP flake on hydrophilic SiO₂ after **(a)** 41 min in ambient and **(b)** 53 min in ambient. Degradation bubbles exist for both time points, but no significant morphological changes are observed. AFM height images for the same region of a ~76.8 nm flake on hydrophobic OTS/SiO₂ after **(c)** 58 min and **(d)** 73 min. Degradation bubbles change shape and coarsen for BP/OTS/SiO₂, even in 15 min. Degradation kinetics are slower for BP on hydrophilic SiO₂ compared to hydrophobic OTS/SiO₂. Green arrows denote the same region for both samples.

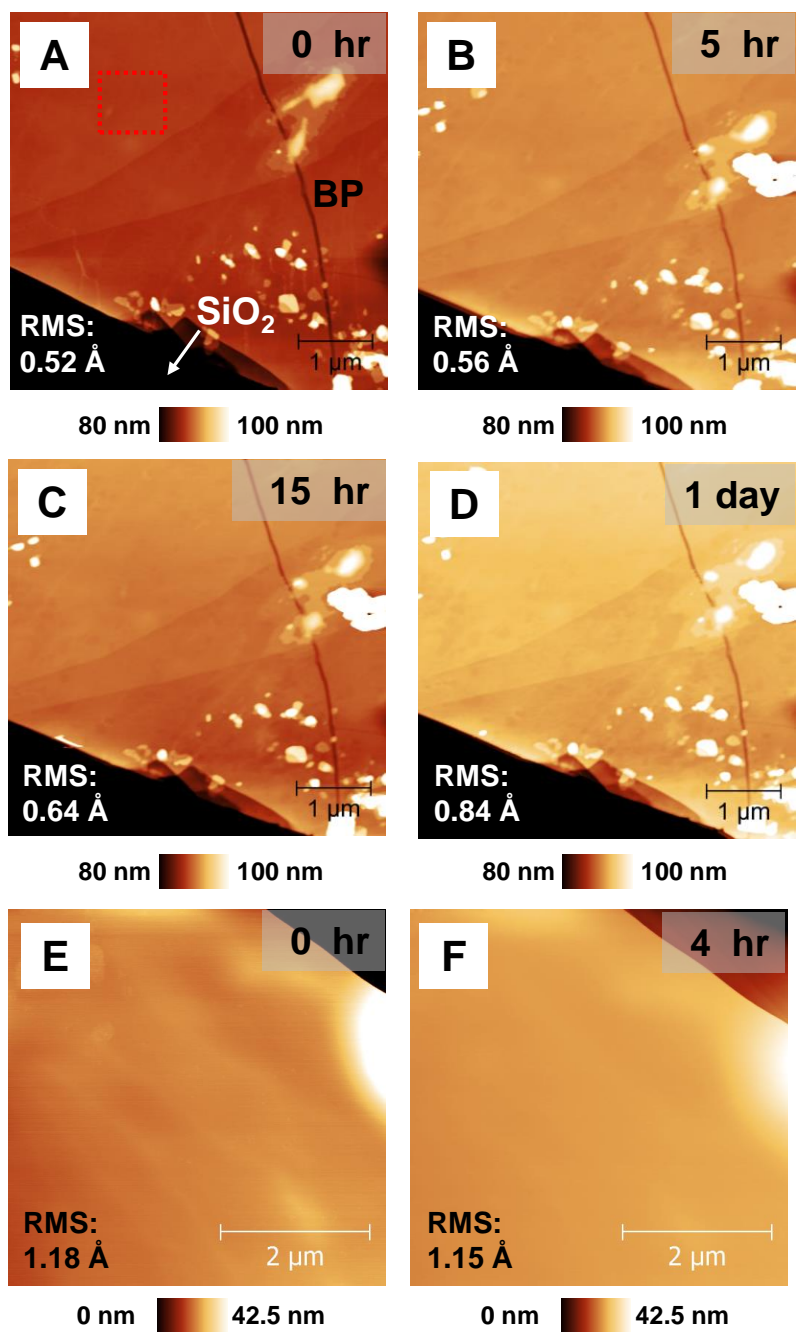

**Figure S13. Preservation of BP in N₂ environment.** AFM height images for a ~86.1 nm thick BP flake on $SiO_2$ **(a)** immediately after insertion in a $N_2$ environmental cell, **(b)** 5 hrs later, **(c)** 15 hrs later, and **(d)** 1 day later. The flake shows no obvious degradation related morphology. AFM height images for a BP flake (thickness less than 100 nm) immediately after placement in **(f)** a $N_2$ environmental cell and **(g)** 4 hrs later. Again, after 4 hrs in $N_2$, no degradation bubbles are evident.

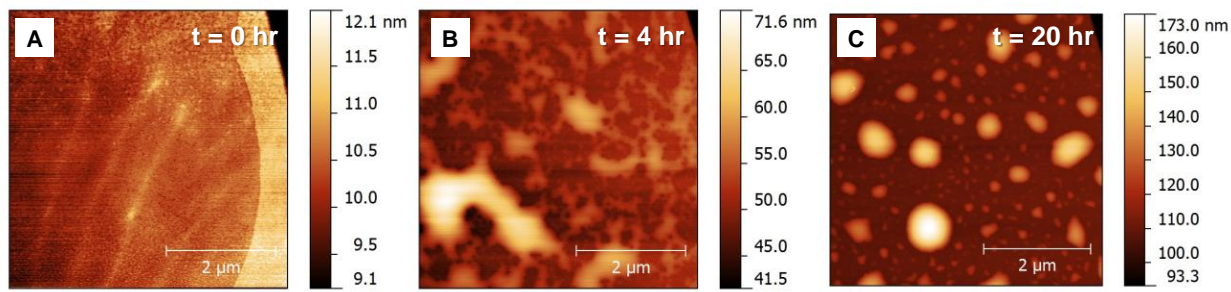

**Figure S14. Degradation of BP in a humid environment.** **(a)** Initial AFM height image of a BP flake on SiO$_2$, taken in a humid environment. **(b)** Same flake after 4 hr. **(c)** Same flake after 20 hr. The degradation bubbles spread and coarsen with time. The supporting movie SM1 provides additional time points.

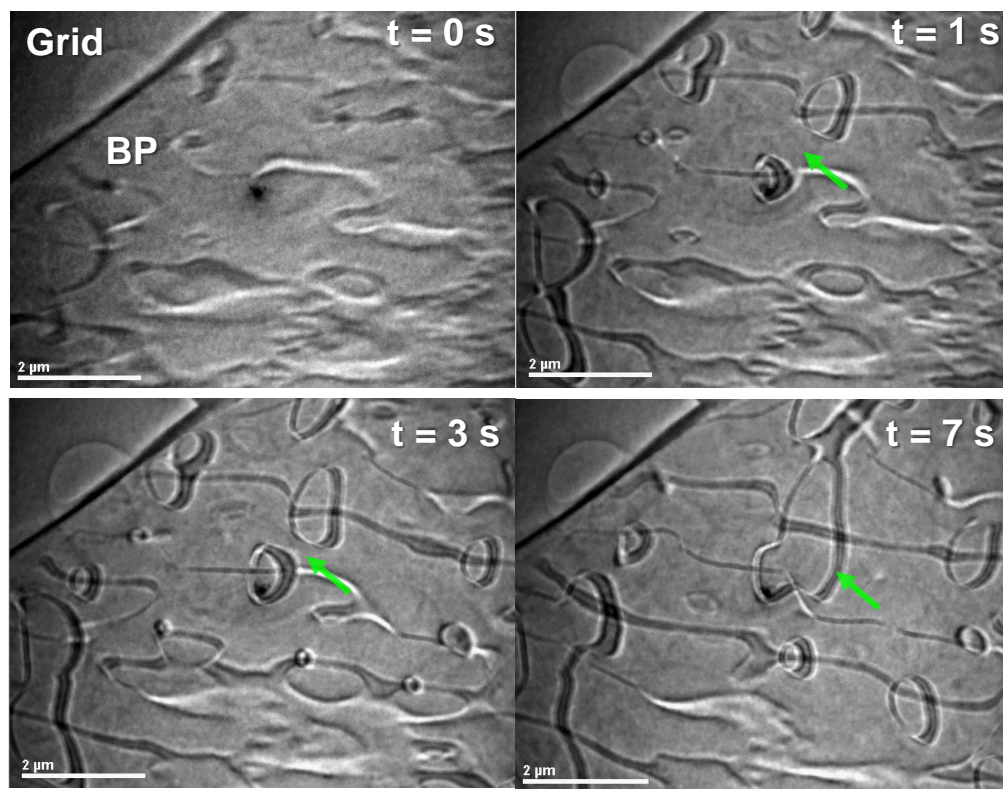

**Figure S15. TEM imaging of BP encapsulated species.** TEM bright field images of a transferred BP flake on Quantifoil, showing trapped species at t = 0, 1, 3, and 7 s. TEM imaging induces movement of the trapped species in the BP (green arrows) over the time scale of seconds. The supporting movie SM2 provides additional time points.

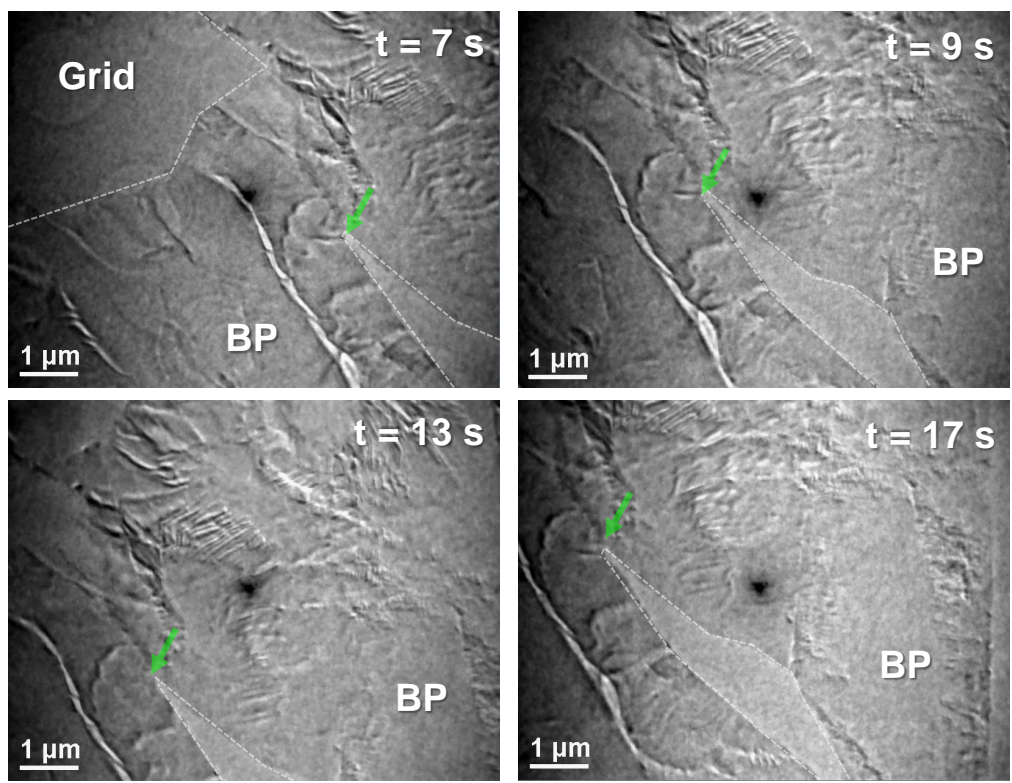

**Figure S16. TEM imaging of BP without any encapsulated species.** TEM bright field images of a transferred BP flake on Quantifoil at t = 7, 9, 13, and 17 s. The green arrow denotes the same position on the flake. Flake was stored in a $N_2$ glove box immediately after flake transfer and before TEM imaging, thereby suppressing ambient degradation. Images are snapshots from the supporting movie SM3, and the time steps are the times listed in SM3. Compared to Fig. S15 and SM2, the flake demonstrates no encapsulated species and is unmodified by subsequent TEM imaging. The lack of damage from continued beam exposure is also supported by the high resolution TEM imaging given in SM4, since that movie was taken after SM3.

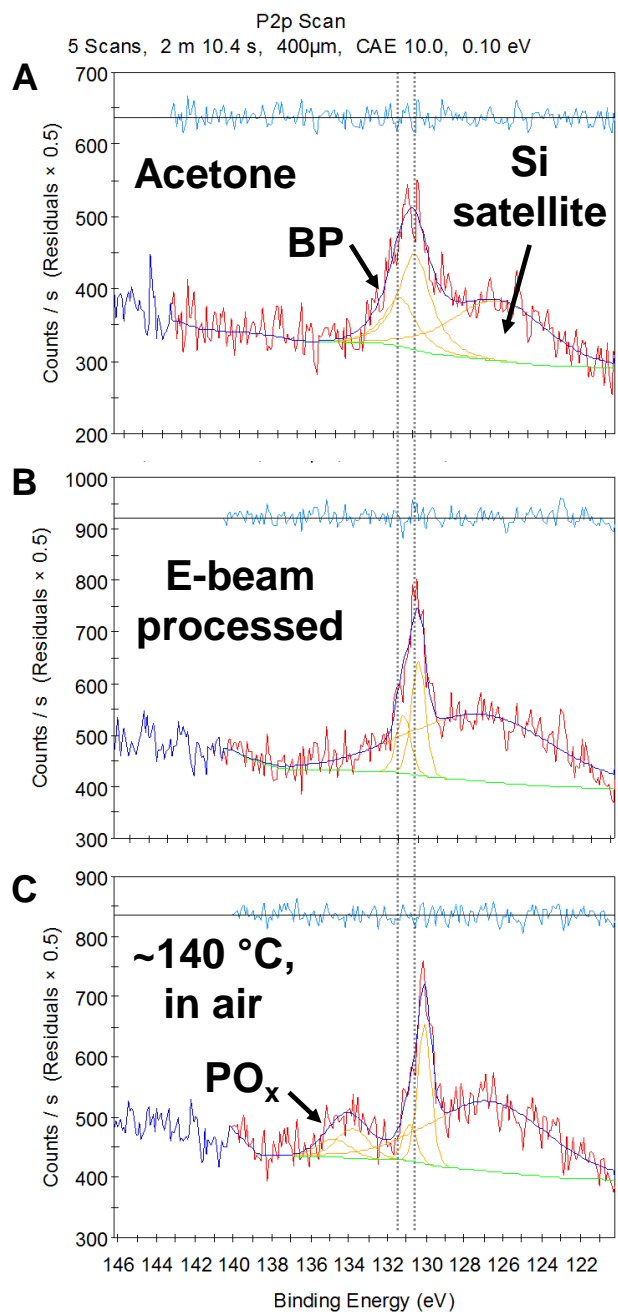

**Figure S17. XPS spectra for BP/SiO₂ exposed to different processing conditions. (a)** P 2p core level spectrum for BP soaked in hydrated acetone for 30 min. Note that this soaking was performed in an ambient environment. BP doublet is apparent, and no oxygenated phosphates are evident. **(b)** P 2p spectrum for BP coated with PMMA, heated at ~140 °C for 90 s, soaked in 3:1 MIBK:IPA for 20 s, and finally soaked in dry acetone (glove box conditions) for 30 min. These conditions mimic those during electron beam lithography (EBL). No oxygenated phosphates (POₓ) are apparent. **(c)** P 2p spectrum for BP exposed to ~140 °C heating in ambient for 90 s. Compared with (b), the BP flakes are not protected by PMMA. A significant POₓ subpeak appears in the spectrum, suggesting BP degradation under these conditions.

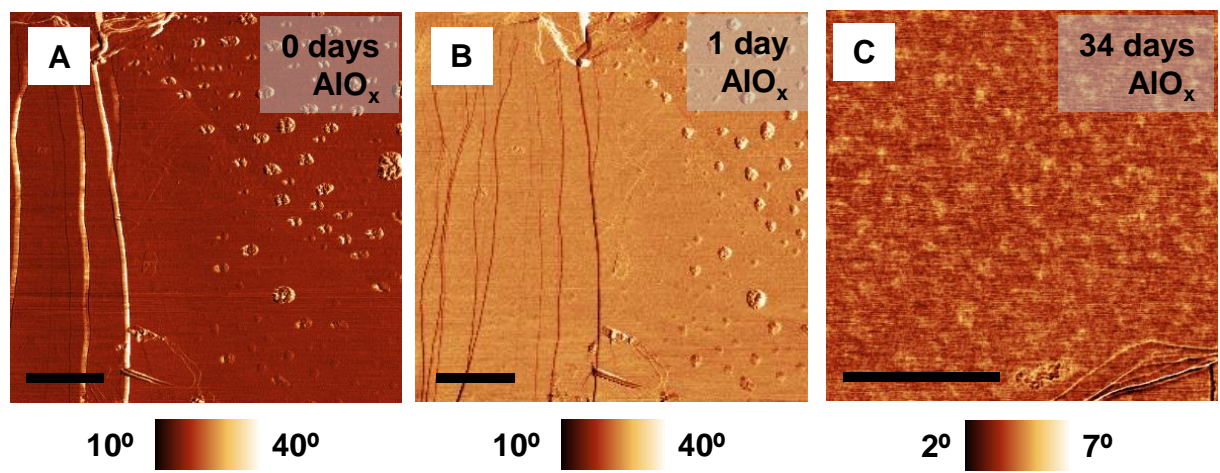

**Figure S18. AFM phase images for AlOₓ passivated BP.** Phase images for a ~72.5 nm thick BP flake covered by ~3 nm ALD AlOₓ **(a)** after ALD and **(b)** 1 day later. No bubbles are present, indicating suppressed BP degradation (the observed features are tape residues that are present even at t = 0). **(c)** Phase images for a different, passivated BP flake (~173.2 nm thickness), having no apparent degradation after 34 days of ambient exposure. Scale bars are 1 μm.

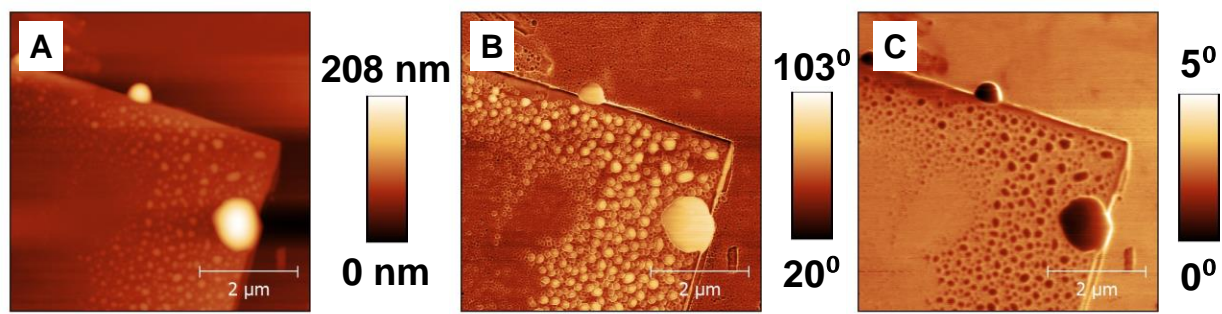

**Figure S19. Electrostatic force microscopy (EFM) images of exfoliated BP on Au.** Normal AFM height **(a)** and phase **(b)** images for a ~31.5 nm thick BP flake on Au. Degradation bubbles are evident after 12 hrs in ambient. **(c)** EFM phase image for the same region in (a) and (b). The EFM phase is lower in the bubble protrusions than in the center of the flake. This EFM contrast is consistent with a lower electrical conductivity in the degradation regions, relative to the rest of the BP flake.

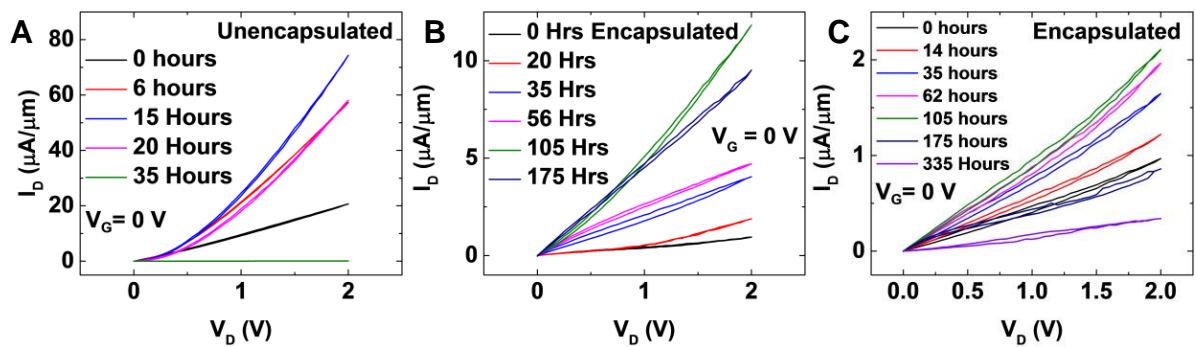

**Figure S20. *I–V* curves for encapsulated and unencapsulated devices.** *I–V* curves for a typical example of (**a**) an encapsulated, (**b**) an unencapsulated device with Ti/Au (Au on top) contacts, and (**c**) an encapsulated Ni/Au (Au on top) device. All device data are at $V_G = 0$, showing transport in the linear regime. Note that the multiple curves represent data collected at discrete times, not discrete gate biases.

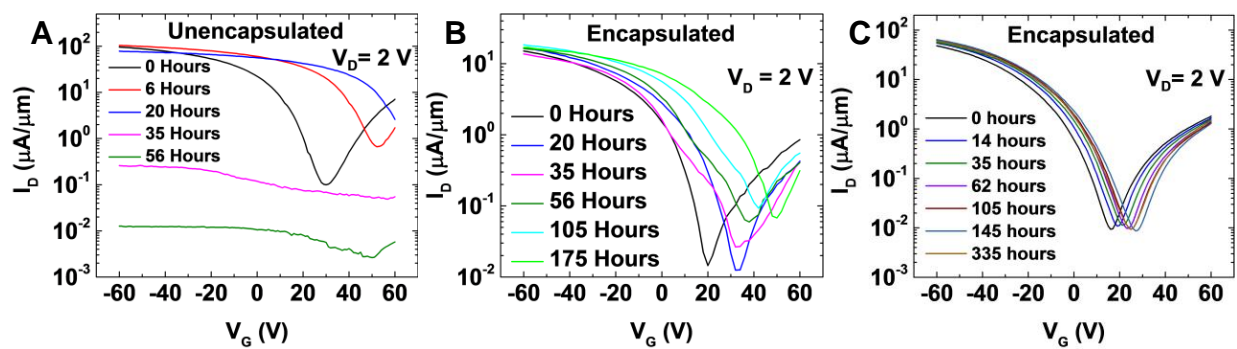

**Figure S21. Log scale transfer curves of encapsulated and unencapsulated devices**. Log scale transfer curve for typical examples of (**a**) an unencapsulated device, (**b**) an encapsulated Ti/Au device, and (**c**) an encapsulated Ni/Au device. Transconductance decays monotonically for the unencapsulated device, while it is preserved for the investigated time period in the encapsulated devices.

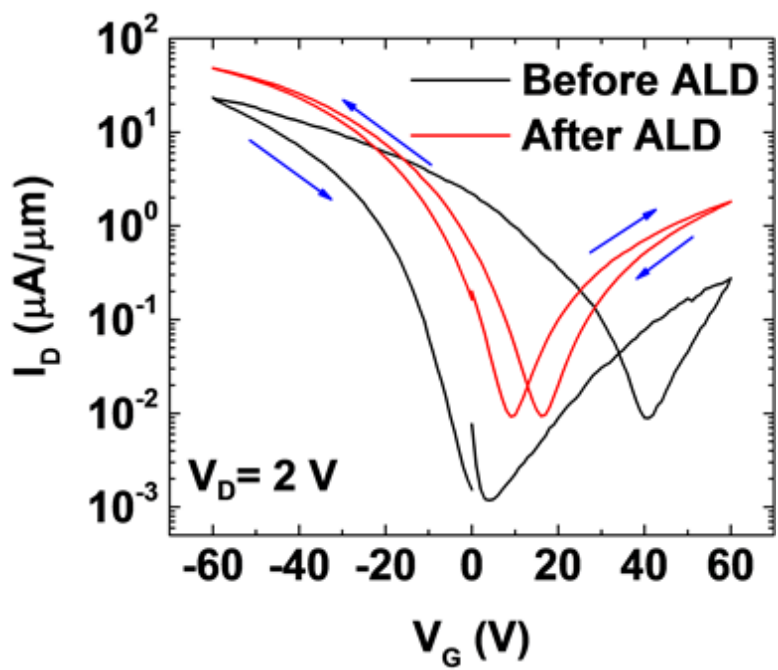

**Figure S22. BP FET transfer curve post ALD.** Transfer curves of a typical Ni/Au device before and after ALD. These curves show an increase in mobility and $I_{ON}/I_{OFF}$ ratio and decrease in threshold voltage and hysteresis. The voltage sweep direction is shown by the green arrow.

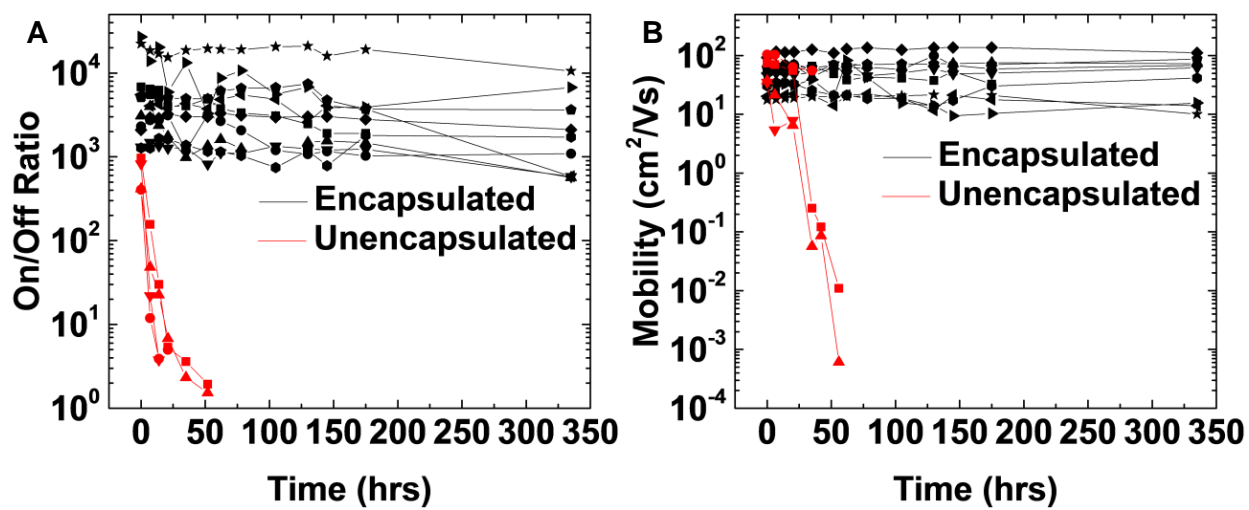

**Figure S23. Comparison of encapsulated Ni/Au and unencapsulated Ti/Au devices.** In unencapsulated Ti/Au devices, the **(a)** $I_{ON}/I_{OFF}$ ratio and **(b)** mobility drop rapidly. Conversely, in the encapsulated Ni/Au devices, these device characteristics are preserved for the two weeks investigated.